\documentclass[11pt,a4paper]{article}
\usepackage{pub}
\usepackage[T1]{fontenc}
\usepackage[english]{babel}
\usepackage{lmodern}
\usepackage{float}
\usepackage{tikz-feynman}
\usepackage{bbm}
\usepackage{bm}
\usepackage{color}
\usepackage{slashed}
\usepackage{graphicx}
\usepackage{braket}
\usepackage{bbold}
\usepackage{subfigure}
\usepackage{subcaption}
\usepackage{dsfont}
\usepackage{todonotes}

\DeclareGraphicsExtensions{.pdf,.png,.jpg}

\allowdisplaybreaks[1]
\def\simg{{\ \lower-1.2pt\vbox{\hbox{\rlap{$>$}\lower6pt\vbox{\hbox{$\sim$}}}}\ }}
\def\siml{{\ \lower-1.2pt\vbox{\hbox{\rlap{$<$}\lower6pt\vbox{\hbox{$\sim$}}}}\ }}

\makeatletter \@addtoreset{equation}{section} \makeatother
\renewcommand{\theequation}{\arabic{section}.\arabic{equation}}

\newcommand{\rmi}[1]{{\mbox{\scriptsize #1}}}
\newcommand{\rmii}[1]{{\mbox{\tiny\rm{#1}}}}

\newcommand{\vrel}{v_\rmi{rel}}

\newcommand{\sw}{s_\rmi{w}}

\begin{document}

\flushbottom

\begin{titlepage}
\begin{centering}

\vfill


{\Large{\bf
Probing Leptophilic Scalar Dark Matter 
\\
with SMEFT and Direct Detection
}} 

\vspace{0.8cm}

S.~Biondini$^a$\footnote{Email: simone.biondini@physik.uni-freiburg.de}, L.~Tiberi$^{b,c,}$\footnote{Email: lorenzo.tiberi@dottorandi.unipg.it}, C.~M.~Bolloni$^{b,c,}$\footnote{Email: chiaramegan.bolloni@studenti.unipg.it} and O.~Panella$^{c,}$\footnote{Email: orlando.panella@pg.infn.it}

\vspace{0.8cm}

$^{a}${\em Institute of Physics, University of Freiburg,
\\
Hermann-Herder-Stra{\ss}e 3, 79014 Freiburg, Germany} 
\\
\vspace{0.15 cm}
$^{b}${\em Dipartimento di Fisica e Geologia, Università degli studi di Perugia
Via A. Pascoli, I-06123, Perugia, Italy
}
\\
\vspace{0.15 cm}
$^{c}${\em INFN, Sezione di Perugia,
\\
Via A. Pascoli, I-06123, Perugia, Italy}

\vspace*{0.8cm}

\end{centering}

\vspace*{0.3cm}

\noindent
\textbf{Abstract}: 
Heavy dark sectors may evade direct production at current colliders while leaving measurable imprints in precision observables and dark matter searches. We study leptophilic scalar dark matter within the Standard Model Effective Field Theory (SMEFT), considering both real and complex scalar dark matter coupled to vector-like leptonic mediators. We perform the complete one-loop matching of the model onto the dimension-six Warsaw basis and combine the resulting precision constraints with predictions for direct and indirect detection. We compute the relic abundance including coannihilations with the mediator, Sommerfeld enhancement, bound-state formation, and one-loop corrections to scalar dark matter annihilation. We also derive the effective dark matter-photon interactions relevant for nuclear recoils, retaining the full dependence on the dark matter and mediator masses and extending previous results beyond the heavy-mediator limit. We find that global SMEFT fits provide competitive constraints for dark matter masses up to approximately $850$ GeV, while direct detection becomes particularly powerful for compressed spectra. The complex scalar scenario is strongly constrained by the charge-radius interaction, whereas real scalar dark matter remains viable across a substantial region of parameter space. Finally, we show that a modified cosmological expansion history can significantly shift the relic-density-compatible parameter space, thereby changing the complementarity among the different probes.

\vfill
\end{titlepage}

\clearpage
\pagenumbering{gobble} 
\tableofcontents

\clearpage
\pagenumbering{arabic}
\setcounter{page}{1}

\section{Introduction}
\label{sec:intro}

The search for the non-luminous and non-baryonic matter in the Universe has challenged the cosmology and particle physics communities for several decades. It is perhaps remarkable how successfully dark matter (DM) hides from us. On the one hand, it reveals its presence through a wealth of gravitational signatures, ranging from galaxy rotation curves and galaxy-cluster dynamics to gravitational lensing. However, it has so far remained elusive to every non-gravitational probe. Establishing such an interaction would not only reveal the particle nature of dark matter, but also provide a decisive step toward understanding its microscopic origin. Although alternative explanations exist, the hypothesis that dark matter consists of one or more new elementary particles remains the simplest and most widely accepted framework in both astrophysics and particle physics \cite{Bertone:2016nfn,Cirelli:2024ssz}.

The appeal and simplicity of the weakly interacting massive particle (WIMP) paradigm have motivated an extensive experimental program searching for dark matter at and above the electroweak scale. This effort encompasses direct-detection experiments, indirect searches for dark matter annihilation or decay products, and collider probes. Despite their remarkable sensitivity, these searches have so far come up empty-handed, progressively excluding large regions of the parameter space of conventional WIMP scenarios \cite{Schumann:2019eaa,Arcadi:2017kky,Cirelli:2024ssz}. In particular, for dark matter candidates carrying electroweak quantum numbers, the interplay between precision theoretical predictions and increasingly sensitive experimental searches has pushed the classic WIMP into narrow corners of the cosmologically viable parameter space, while some of the simplest realizations have been excluded altogether. Nevertheless, it would be premature to conclude that the WIMP paradigm has been exhausted. The particle nature of dark matter still admits a broad landscape of theoretically well-motivated possibilities, which remain experimentally viable. In particular, if the dark sector lies well above the electroweak scale, it may evade direct production and detection at current experiments.

This observation motivates a complementary strategy. Rather than searching for dark matter particles themselves, one may seek the imprints left by a heavy dark sector in precision observables through quantum corrections. Assuming that the dark matter particle and its associated states are heavier than the electroweak scale, their low-energy effects can be systematically encoded in the Standard Model Effective Field Theory (SMEFT) \cite{Buchmuller:1985jz,Grzadkowski:2010es,Brivio:2017vri}, which provides a model-independent description of heavy new physics. Recent advances in global SMEFT analyses \cite{Bartocci:2023nvp,Bartocci:2024fmm}, driven by increasingly precise measurements and improved theoretical predictions, have considerably enhanced, and made more robust, the sensitivity to new physics at the multi-TeV scale. In this framework, one can match a specific dark matter model onto the SMEFT and translate constraints on effective operators into bounds on the underlying parameters of the dark sector, such as masses and couplings. Precision measurements therefore offer a powerful and genuinely complementary avenue to explore heavy dark sectors, extending the reach of conventional dark matter searches well beyond their direct kinematic limits.

The interplay between heavy dark sectors and SMEFT has recently attracted growing attention~\cite{Freitas:2014jla,Cepedello:2022pyx,Cepedello:2023yao,Gargalionis:2024jaw,Olgoso:2025jot,Biondini:2025gpg,Biondini:2026ryb}, leading to systematic classifications of renormalizable dark sectors and their low-energy phenomenology. In particular, models in which dark matter couples to Standard Model fermions through renormalizable mediators provide a well-defined framework in which precision observables, collider searches, and dark matter experiments can be investigated on equal footing. This broad class of models is often referred to in the literature as $t$-channel dark matter models~\cite{Arina:2020udz,Arina:2025zpi,Garny:2015wea}, although different naming conventions are also adopted, such as dark matter interacting with a charged mediator \cite{Ibarra:2015fqa} or simply lepton-portal dark matter \cite{Bai:2014osa}. Here the dark matter particle is accompanied by a mediator carrying the appropriate Standard Model gauge quantum numbers, thereby realizing a gauge-invariant and renormalizable interaction. A discrete $\mathcal{Z}_2$ symmetry, under which \emph{both} the DM and the mediator are odd, stabilizes the DM particle and, at the same time, implies that the Wilson coefficients of the corresponding dimension-six SMEFT operators are generated for the first time at one loop~\cite{Cepedello:2022pyx,Cepedello:2023yao,Gargalionis:2024jaw,Guedes:2024vuf}. The matching onto SMEFT is therefore technically non-trivial, and we employ automated functional methods~\cite{Fuentes-Martin:2022jrf} to derive the Wilson coefficients.

For couplings between dark matter and Standard Model particles of order unity, the observed dark matter abundance is naturally obtained through thermal freeze-out. Furthermore, when the dark matter particle and the mediator are moderately degenerate in mass, the relic abundance is affected by coannihilation processes \cite{Griest:1990kh,Edsjo:1997bg}, requiring the mediator to be included as an active degree of freedom in the freeze-out dynamics. This compressed spectrum, which is consistent with a \emph{one-step} matching onto the SMEFT, makes the treatment of near-threshold effects in mediator pair annihilations, such as the Sommerfeld enhancement \cite{Hisano:2004ds,Iengo:2009ni,Cassel:2009wt} and bound-state formation \cite{Feng:2009mn,vonHarling:2014kha}, an essential ingredient of a consistent determination of the dark matter relic abundance in such DM models. The impact of near-threshold effects for $t$-channel models with leptophilic mediators has been studied earlier in Refs.~\cite{Bollig:2021psb,Biondini:2022ggt,Biondini:2025gpg}.

In Ref.~\cite{Biondini:2025gpg}, we investigated a leptophilic fermionic dark matter model interacting through a scalar mediator coupled to both left- and right-handed leptons, establishing its SMEFT description and exploring the interplay between precision observables, the relic abundance, and direct-detection constraints. Nuclear-recoil signals, and consequently direct-detection limits, are relevant \emph{even} for leptophilic dark matter models, despite arising only at the one-loop level~\cite{Kopp:2014tsa,Ibarra:2015fqa,Weiner:2012cb,Kavanagh:2018xeh}. Their loop-induced nature renders their calculation considerably more involved than in conventional WIMP scenarios. In particular, direct-detection constraints are especially stringent for Dirac dark matter \cite{Ibarra:2015fqa,Kopp:2014tsa,Biondini:2025gpg,Biondini:2026ryb}, while they remain relevant also for Majorana dark matter through the loop-induced anapole interaction. Our previous analysis consistently incorporated these loop-induced direct-detection effects in the phenomenological study. Such effects have not always been included in phenomenological analyses of leptophilic dark matter models \cite{Olgoso:2025jot,Cepedello:2023yao}, despite their potential impact on the allowed parameter space.

In the present work we turn to the complementary class of leptophilic scalar dark matter models, in which a real or complex scalar communicates with the SM through a vector-like fermion. Besides completing the study of the simplest renormalizable leptophilic $t$-channel dark matter models within the SMEFT framework, this scenario allows us to considerably extend the phenomenological analysis. In particular, the loop-induced nature of both the SMEFT matching and the direct-detection observables calls for a consistent treatment of precision constraints and dark matter searches within a common framework. We therefore perform the complete one-loop matching onto the dimension-six SMEFT operators, revisit the thermal freeze-out including coannihilation and near-threshold effects, and considerably improve the treatment of direct detection. For the present model, only approximate expressions for the dark matter--nucleon interactions are available in the limit $m_{\rmi{med}}\gg m_{\rmii{DM}}$~\cite{Kavanagh:2018xeh}, which is not representative of the moderate and compressed mass spectrum accommodating the cosmological DM abundance and required to consistently perform the one-step matching onto the SMEFT. We therefore carry out a dedicated calculation of the complete one-loop dark matter--photon effective interactions with full dependence on the dark matter and mediator masses, extending previous results beyond the heavy-mediator limit. This enables a comprehensive assessment of the complementarity between precision observables, direct searches, and the cosmologically viable parameter space of leptophilic scalar dark matter.

The remainder of this paper is organized as follows. In Sec.~\ref{sec:model} we introduce the dark matter model and specify the assumptions adopted throughout our analysis. Sec.~\ref{sec:SMEFT_and_DM} presents the matching onto the SMEFT and discusses the Wilson coefficients relevant for constraining the model parameter space. The determination of the dark matter relic abundance is described in Sec.~\ref{sec:relic}, while the experimental probes, with particular emphasis on direct detection, are discussed in Sec.~\ref{sec:constraints}. In Sec.~\ref{sec:results} we combine the constraints from precision observables, direct searches, and the relic-density requirement to delineate the viable parameter space of the model. Finally, our conclusions are presented in Sec.~\ref{sec:conclusions}. Additional technical details and supplementary materials are collected in the appendices \ref{app:WC_scalar_DM}-\ref{app:cross_sections_relic}.

\section{Leptophilic dark matter}
\label{sec:model}

In this section we briefly describe the dark sector considered throughout this work. We study a renormalizable realization of the broad class of leptophilic dark matter models, sometimes referred to in the literature as $t$-channel dark matter models. The dark matter particle is assumed to be a Standard Model (SM) gauge singlet and to couple to a SM fermion. Gauge invariance then requires the introduction of a mediator whose SM quantum numbers are uniquely determined once the spin of the DM particle and the SM fermion are specified. The DM particle and the mediator are taken to be odd under a $\mathcal{Z}_2$ symmetry, whereas all SM fields are even. Together with the kinematic condition $m_{\rmii{DM}} < m_{\rmi{med}} + m_f$, the stability of the DM particle is guaranteed.

In this work we focus on scalar DM, for which the mediator is necessarily a fermion. We further restrict our attention to couplings with right-handed charged leptons, corresponding to the simplest electroweak charge assignment, cfr.~Eq.~\eqref{eq:portal_scalar_DM}. Since the DM particle carries no SM gauge charges and does not interact with quarks, its interactions with nucleons arise only \emph{radiatively}, with the effective interactions relevant for nuclear recoils being generated at one-loop order, see e.g.~\cite{Kopp:2014tsa,Ibarra:2015fqa}. Likewise, the leading SMEFT Wilson coefficients induced by the dark sector are also generated at one loop. The two classes of observables are therefore sensitive to quantum effects at the same perturbative order, making it essential to treat both consistently in a comprehensive phenomenological analysis. 
The present work also complements our earlier study of fermionic dark matter~\cite{Biondini:2025gpg}, while substantially extending the phenomenological analysis through a complete treatment of the loop-induced dark matter--nucleon scattering amplitudes.

We denote the dark matter field by $S$, the vector-like fermion mediator by $F^a$, and the right-handed SM charged lepton by $e^a$, where $a=e,\mu,\tau$ labels the lepton flavour. The beyond-the-Standard-Model (BSM) Lagrangian reads
\begin{equation}
\mathcal{L}_{\rmii{BSM}}
=
\mathcal{L}_{\rmii{SM}}
+
\mathcal{L}_{S}
+
\mathcal{L}_{F}
+
\mathcal{L}_{\rmi{portal}} \, ,
\label{eq:lag_schem}
\end{equation}
where $\mathcal{L}_{\rmii{SM}}$ denotes the SM Lagrangian, while $\mathcal{L}_{S}$ and $\mathcal{L}_{F}$ describe the scalar DM and the vector-like fermion mediator, respectively.

For a real scalar DM particle,
\begin{align}
    \mathcal{L}_{S}
    =
    \frac{1}{2}\partial^\mu S\,\partial_\mu S
    -
    \frac{m_{S,0}^2}{2}S^2
    -
    \frac{\lambda_S}{4!}S^4 \, .
\end{align}
The extension to complex scalar DM is straightforward, and we shall explicitly point out the differences between the real and complex cases whenever relevant. The parameter $m_{S,0}$ denotes the DM mass before electroweak symmetry breaking, since the physical mass generally receives an additional contribution from the Higgs vacuum expectation value through the Higgs portal, cf.~Eq.~\eqref{eq:portal_scalar_DM}.

The Lagrangian of the vector-like fermion mediator and its covariant derivative read
\begin{equation}
    \mathcal{L}_{F}
    =
    \bar{F}^a\left(i\slashed{D}-m_F\right)F^a \, , \quad D_\mu=\partial_\mu-i g_1 Y_F B_\mu \, ,
\end{equation}
with $g_1$ the SM hypercharge gauge coupling and $Y_F$ the mediator hypercharge. After electroweak symmetry breaking, the mediator couples to both the photon and the $Z$ boson.

The portal Lagrangian describes the interaction between the dark and visible sectors. In full generality it contains both the Yukawa interaction and the scalar Higgs portal,
\begin{equation}
\mathcal{L}_{\rmi{portal}}
=
-
y\,S\,\overline{F}^a P_R e^a
+
\text{h.c.}
+
\frac{\lambda_{SH}}{2}H^\dagger HS^2 \, .
\label{eq:portal_scalar_DM}
\end{equation}
Here one sees that the mediator hypercharge is then fixed to $Y_F=Y_{e_R}=-1$.\footnote{The Gell-Mann--Nishijima relation is taken in the form $Q_{\rmi{em}} = I_{W}+Y$.} Throughout this work we assume the Higgs-portal coupling to be negligible, $\lambda_{SH}\ll1$, and focus on the Yukawa interaction. This choice isolates the phenomenology associated with the leptophilic interaction and avoids the interplay with the well-studied Higgs-portal phenomenology of scalar singlet dark matter. This is also the common assumption adopted in the phenomenology of $t$-channel DM models \cite{Garny:2015wea,Arina:2020udz,Arina:2020tuw,Arina:2025zpi}. For a representative work on the interplay between the Higgs portal and the Yukawa-like interaction with leptons see \cite{Ibarra:2014qma}, whereas for scalar DM interacting with quarks see e.g.~\cite{Biondini:2019int}. 

Let us briefly comment on the flavour structure adopted in our work. Our main objective is to confront our model with global SMEFT analyses, which are only feasible under assumptions about the flavour structure of the operators. In our work, we adopt the recent analyses in Refs.~\cite{Bartocci:2023nvp,Bartocci:2024fmm}, which exploit the framework of Minimal Flavour Violation (MFV), and formulate the DM model accordingly, cf.~\eqref{eq:portal_scalar_DM}. The underlying flavour symmetry,
\begin{equation}
\mathcal{G}_{\rm flavor}=U(3)^5\, ,
\end{equation}
is assumed to be broken exclusively by the SM Yukawa interactions
\cite{Gerard:1982mm,Chivukula:1987py,Hall:1990ac,DAmbrosio:2002vsn}. This hypothesis drastically reduces the number of independent flavour structures in the effective theory, allowing for a meaningful global SMEFT analysis while remaining consistent with the stringent experimental constraints on flavour-changing processes. 
Other recent global fits adopting the same flavour assumptions can be found in Refs.~\cite{deBlas:2025xhe,Giani:2023gfq}. Nevertheless, we rely on Ref.~\cite{Bartocci:2024fmm} both for consistency with our previous work \cite{Biondini:2025gpg} and because it provides the strongest constraints in our exclusion plots in Sec.~\ref{sec:results}. 

In line with these flavour assumptions, the vector-like fermion mediator transforms in the same flavour representation as the right-handed charged leptons, namely as a triplet under $U(3)_e$. The dark sector therefore contains three degenerate copies of the mediator. We further assume the Yukawa couplings to be real, so that no additional source of CP violation beyond the SM is introduced. Under these assumptions, the Yukawa interaction is flavour universal,
$y^e=y^\mu=y^\tau\equiv y$.
The interplay between leptophilic and quarkphilic dark matter models and more general flavour constructions has been recently discussed in Refs. \cite{Belfatto:2025ids,Arina:2025zpi,Biondini:2026ryb}.

\section{Matching to SMEFT}
\label{sec:SMEFT_and_DM}

In this section we derive the matching of the leptophilic dark matter model introduced in Sec.~\ref{sec:model} onto SMEFT. Since the dark sector is assumed to lie above the electroweak scale, its low-energy effects can systematically be described in terms of higher-dimensional operators involving only SM fields. Throughout this work we retain operators up to dimension six and adopt the Warsaw basis of Ref.~\cite{Grzadkowski:2010es}, and take $m_F > m_S > 250$ GeV. The largest scale in the model, namely $m_F$, is taken to be the high-energy scale through which the expansion of the SMEFT operators is organized.\footnote{Typical choices for the electroweak scale in SMEFT are $E_{\rmii{EW}}=v_h$ or $E_{\rmii{EW}}=m_Z$. Global fits for leptonic
observables are extracted from LEP data at energies up to 200 GeV. Our choice of the lowest DM mass scale
aligns with earlier works, see e.g.~\cite{Freitas:2014jla} and \cite{Cepedello:2023yao}, where the dark sector particle masses are taken
to be as small as 200 GeV. Even if such DM masses lie at the edge of the validity of the EFT, the SMEFT
constraints may still be indicative for such moderately small masses.} 

The matching is performed by integrating out the heavy degrees of freedom at a common scale. This one-step procedure is justified when the DM particle and the mediator are sufficiently close in mass. We therefore consider both moderately compressed and nearly degenerate  mass spectra in the dark sector,\footnote{For larger mass splittings, a one-step matching is no longer appropriate. In this case, the mediator should first be integrated out at the scale $m_F$, followed by renormalization-group evolution down to the dark matter scale $m_S$, where the DM field is subsequently integrated out.}
\begin{equation}
m_F = m_S(1+\delta m)\, , \qquad
0.005 \leq \delta m \leq 1.0 \, .
\end{equation}
The relative mass splitting is defined as $\delta m = \Delta m / m_S$, where $\Delta m = m_F-m_S$.
This setup allows us to match the ultraviolet theory directly onto SMEFT and to make use of existing global analyses of the Wilson coefficients in the Warsaw basis. In particular, consistently with the MFV hypothesis adopted in Sec.~\ref{sec:model}, we employ the global fit of Ref.~\cite{Bartocci:2024fmm}, which takes into account NLO observables and renormalization group evolution effects. 

The matching is carried out using the functional approach of Ref.~\cite{Cohen:2020fcu}, as implemented in the automated package \texttt{Matchete}~\cite{Fuentes-Martin:2022jrf}. Owing to the exact $\mathcal{Z}_2$ symmetry of the ultraviolet theory, the relevant dimension-six operators are not generated at tree level. Consequently, the corresponding Wilson coefficients arise for the first time at one-loop order. 

We truncate the effective theory at dimension six throughout the analysis, consistently with the global SMEFT fit of Ref.~\cite{Bartocci:2024fmm}. Higher-dimensional operators are suppressed by additional powers of the heavy scale and provide subleading corrections in the regime where the effective-field-theory expansion is valid. 



\subsection{Four-fermion operators in the Warsaw basis}
\label{subsec:four_fermion_matching}

Among the operators generated by the one-loop matching, the four-fermion operators provide the most relevant constraints on the parameter space considered in this work. Their Wilson coefficients contain both pure-gauge contributions and terms involving the dark-sector Yukawa coupling, with a non-trivial dependence on the DM--mediator mass ratio. We therefore discuss this class explicitly and collect below the coefficients that are relevant for our phenomenological analysis. Defining the ratio between the squared masses of dark matter and the mediator as
\begin{equation}
    \rho \equiv \frac{m^2_S}{m_F^2} \, ,
\end{equation}
we find, for the fully leptonic sector, the following results 
\begin{align}
C_{ee}^{(S^* = S)}
    &=
    \frac{2}{(4\pi)^2 m_F^2}
    \left(
        -\frac{1}{5} g_1^4 + \frac{1}{12}g_1^2 y^2 \mathcal{F}_1(\rho)
    \right) \, ,
    \label{Cee_scalar_real}
    \\
    C_{ee}^{(S^* \neq S)}
    &=
    \frac{2}{(4\pi)^2 m_F^2}
    \left(
        -\frac{1}{5} g_1^4
        + \frac{1}{12}g_1^2 y^2 \mathcal{F}_1(\rho) + \frac{1}{16} y^4 \mathcal{F}_2(\rho) 
    \right) \, ,
     \label{Cee_scalar_complex}
     \\
    C_{le} &= -\frac{1}{(4\pi)^2 m_F^2}
    \left(  -\frac{2}{5} g_1^4 + \frac{1}{12} g_1^2 y^2 \mathcal{F}_1(\rho) \right) \, , \quad
 C_{ll}
    =
    -\frac{g_1^4}{10 (4\pi)^2 m_F^2}
     \, ,   \label{Cle_Cll}
\end{align}
whereas the mixed lepton-down-type quark operator is 
\begin{align}
    C_{ed}
    &=
   \frac{1}{(4\pi)^2 m_F^2} \left( - \frac{4}{15} g_1^4 + \frac{1}{18} y^2 g_1^2 \mathcal{F}_1(\rho)\right) \, .
\end{align}
The loop functions $\mathcal{F}_i(\rho)$, together with the remaining Wilson coefficients generated by the matching, are collected in App.~\ref{app:WC_scalar_DM}. There is a difference between real and complex scalar DM in the expressions of the matching coefficient $C_{ee}$. In the real scalar case the term proportional to $y^4$ is absent.  Indeed, the origin of such contribution, in the complex scalar DM option, originates from the additional contractions associated with a non-self-conjugate scalar field.

The structure of the Wilson coefficients in Eqs.~\eqref{Cee_scalar_real}-\eqref{Cle_Cll} offers a quick overview of how they may constrain the model parameter space. The coefficient $C_{ee}$ contains terms proportional to $g_1^4$,  $\,g_1^2 y^2$, and $y^4$, and therefore exhibits a direct sensitivity to the dark-sector masses, through the loop functions, \emph{and} coupling. The same applies to the coefficients $C_{le}$ and $C_{ed}$.  The latter is shown together with the fully leptonic operators because it constraints the parameter space of the model with the global fit of Ref.\cite{Bartocci:2024fmm}. In contrast, $C_{ll}$ is generated entirely by gauge interactions and is independent of $y$. Their dependence on the mass ratio $\rho$ is therefore relevant in the compressed regime considered here and cannot in general be captured by the limit of a parametrically heavy mediator, $\rho \to 0$.

We confront these coefficients with the global SMEFT analysis of Ref.~\cite{Bartocci:2024fmm}, performed under the MFV assumptions adopted in our model. The four-lepton coefficients $C_{ee}$, $C_{le}$, and $C_{ll}$ are constrained by a combination of low-energy leptonic observables, electroweak precision measurements, and LEP data. The latter reached a maximal centre-of-mass energy of $\sqrt{s}=209~\mathrm{GeV}$~\cite{Electroweak:2003ram}. We therefore apply the constraints from the fully leptonic operators for dark-sector masses satisfying $m_S,m_F \geq 250~\mathrm{GeV}$, for which the characteristic energy scale probed by the fit remains below the mass scale of the states integrated out. The validity of the EFT description improves as the dark-sector mass scale is increased.

The semileptonic coefficient $C_{ed}$ is additionally constrained by observables involving quarks and therefore provides a complementary probe of the parameter space. The dominant constraints on $C_{ed}$ arise from parity-violation experiments (PVE) and high-energy Drell--Yan processes. Throughout the relevant parameter space, the Drell--Yan constraints provide the stronger bound and are based on LHC data extending up to invariant masses of approximately $3~\mathrm{TeV}$~\cite{CMS:2021ctt,ATLAS:2020zms}. We therefore apply the corresponding Drell--Yan constraint to dark-sector states with masses above $3~\mathrm{TeV}$, where the EFT description is appropriate for the energy range probed by these data.

In our numerical analysis we consistently evolve the Wilson coefficients between the matching scale and the scale at which the global fit is provided. The explicit values that are employed in our phenomenological analysis are \cite{Bartocci:2023nvp,Bartocci:2024fmm}
\begin{align}
\frac{C_{ee}(\Lambda)}{\Lambda^2} = \frac{[-0.14, 0.03]}{\textrm{TeV}^2}
\, ,
\quad \frac{C_{l l}(\Lambda)}{\Lambda^2} = \frac{[0.0, 0.21]}{\textrm{TeV}^2} \, , \quad \frac{C_{ed}(\Lambda)}{\Lambda^2} = \frac{[0.0, 1.2]}{\textrm{TeV}^2}
\, .
\label{bounds_relevant_WCs}
\end{align}
Since the Wilson coefficients are generated at the matching scale $\mu \simeq m_F \simeq m_S$, while the global fit is quoted at a fixed reference scale $\Lambda = 4$ TeV, the corresponding renormalization-group evolution must be taken into account when applying the SMEFT constraints.  This approach is more rigorous than simply rescaling the bounds in Eq.~\eqref{bounds_relevant_WCs} as $C_i(m_F) = C_i(\Lambda) \times (m_F/\Lambda)^2$, which neglects potential operator mixing effects. However, as also found in our earlier study \cite{Biondini:2025gpg}, the running induces effects at the level of a few percent across the full range of the mediator mass explored in our work. The quantitative RGE evolution of the matching coefficients has been performed using \texttt{DsixTools}~\cite{Fuentes_Mart_n_2021} and cross-checked with \textit{RGESolver}~\cite{Di_Noi_2023}.

\begin{figure}
    \centering
    \includegraphics[width=0.475\linewidth]{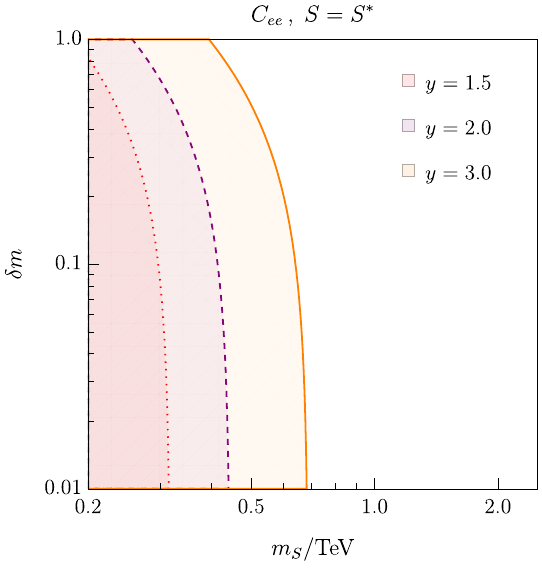}
    \hspace{0.5 cm}
     \includegraphics[width=0.475\linewidth]{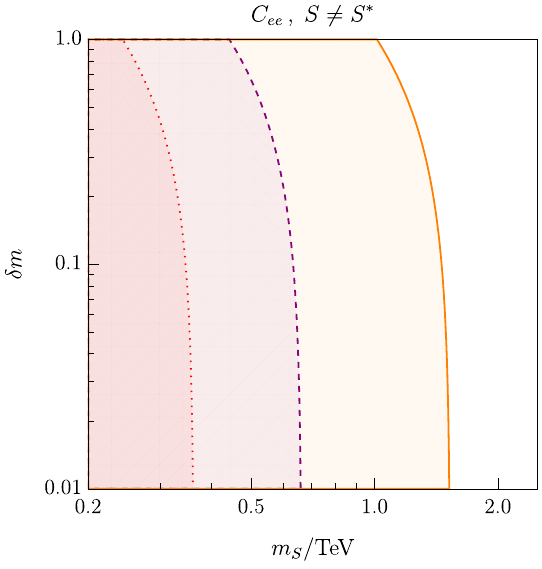}
   \caption{Exclusion regions in the $(m_S, \delta m)$ plane for the matching coefficients $C_{ee}^{S=S^*}$ and $C_{ee}^{S\neq S^*}$, given in Eqs.~\eqref{Cee_scalar_real} and \eqref{Cee_scalar_complex}, respectively, for fixed values of the Yukawa couplings as indicated in the legend.}
    \label{fig:Cee_real_complex}
\end{figure}

In Fig.~\ref{fig:Cee_real_complex} we show the constraints from the matching coefficients $C_{ee}^{S=S^*}$ and $C_{ee}^{S \neq S^*}$ on the model parameter space $(m_S, \delta m)$ for fixed values of the Yukawa coupling. The left panel shows the real scalar dark matter case, whereas the right panel corresponds to the complex scalar case. One can see that the constraints from the complex scalar case are stronger than those from the real scalar case.

\section{Relic density}
\label{sec:relic}
In this section we discuss the main features of the thermal freeze-out for the leptophilic dark matter. We describe the Boltzmann equation upon including coannihilations, and the corresponding near-threshold effects for the mediator in Sec.~\ref{sec:coann_DM}, whereas higher order corrections to the annihilation cross section of the scalar DM and fermion mediator are addressed in Sec.~\ref{sec:higher_order_S}. 
\subsection{Scalar DM with coannihialtions}
\label{sec:coann_DM}

When the dark matter particle and the fermion mediator have masses of the same order, with \(m_F \gtrsim m_S\), the thermal history can receive important contributions from processes involving the mediator. In particular, if the spectrum is sufficiently compressed, these contributions become especially relevant because the mediator remains thermally populated during dark matter freeze-out. For a relative mass splitting \(\delta m \lesssim 0.3\)~\cite{Griest:1990kh,Baker:2015qna}, the mediator abundance is typically not Boltzmann suppressed enough to be neglected, placing the model in the coannihilation regime~\cite{Griest:1990kh,Edsjo:1997bg}. In this regime, the relic abundance is determined by the combined annihilation and coannihilation dynamics of the dark sector, including dark matter pair annihilations, processes involving one dark matter particle and one mediator, and mediator pair annihilations.

More generally, even away from the strongly compressed limit, mediator-related processes can affect the relic abundance whenever the mediator remains sufficiently populated and chemical equilibrium between the dark matter and mediator is efficiently maintained. This is guaranteed by the values of Yukawa couplings considered in our work. The precise importance of coannihilations therefore depends not only on the mass splitting, but also on the relative size of these couplings and on the velocity dependence of the dark matter annihilation cross section. Specifically for the real scalar DM option, where the DM pair annihilation feature a $\vrel^4$ suppression (cfr.~\eqref{SS_ann_tree_level}), the coannihilations play a role already for moderate relative splittings. We consequently consider the general regime \(m_F \gtrsim m_S\), including the moderately compressed and nearly degenerate mass spectrum as a phenomenologically relevant case, and consistently account for the corresponding mediator contributions to the thermal relic abundance.

Under these conditions, the evolution of the dark sector can still be described by a single Boltzmann equation~\cite{Griest:1990kh,Edsjo:1997bg},
\begin{equation}
    \frac{d n}{dt} + 3H n = -\langle \sigma_{\rm eff} v_{\rm rel} \rangle
    \left(n^2-n_{\rm eq}^2\right)\, ,
    \label{BE_start}
\end{equation}
where $n$ denotes the total number density of the dark sector particles, the scalar DM and the mediator, and $H$ the Hubble rate. The dark sector contains three exactly degenerate copies of the fermion mediator, labelled by the flavour index $a=e,\mu,\tau$; accordingly $n_F = \sum_{a} n_{F_a}$. 
Whenever relevant, we shall explicitly comment on the differences between the real and complex scalar scenarios throughout the paper. 

The effective annihilation cross section entering Eq.~\eqref{BE_start} is given by
\begin{equation}
   \sigma_{\rm eff} v_{\rm rel}
   =
   \frac{1}{\left(\sum_k n_k^{\rm eq}\right)^2}
   \sum_{i,j}
   n_i^{\rm eq}
   n_j^{\rm eq}
   \sigma_{ij}v_{\rm rel}\, ,
\end{equation}
where the indices $i,j$ run over all particles belonging to the dark sector. Assuming Maxwell--Boltzmann statistics for the equilibrium number densities of the dark-sector species, and retaining only the leading non-relativistic contribution in the momentum expansion, the effective annihilation cross section can be written in the compact form\footnote{In the numerical determination of the relic abundance, we employ the equilibrium number densities expressed in terms of modified Bessel functions. The approximate expression of Eq.~\eqref{eff_Xsection_coann_real_S} is nevertheless useful to identify the relative importance of the various annihilation channels and to understand their dependence on the mass splitting.}
\begin{align}
     \sigma_{\textrm{eff}} \vrel &= \left[ \sigma_{S S} \vrel +  \frac{4 g_F \,}{g_S} e^{- \frac{\Delta m}{T}}  \left( 1+\delta m \right)^{\frac{3}{2}} \sigma_{S \bar{F}} \vrel  \right.
    \nonumber
    \\
    &\left. +  \frac{ 2 g_F^2}{g_S^2} e^{-2  \frac{\Delta m}{T}} \left( 1+\delta m \right)^{3} \left(   \sigma_{F \bar{F}} \vrel + \sigma_{F F} \vrel\right) \right]
    \frac{n_f}{\left[ 1+ \frac{2  n_f g_F}{g_S} e^{- \frac{\Delta m }{T}} \left( 1+\delta m \right)^{\frac{3}{2}} \right]^2} \, .
    \nonumber
    \\
    \phantom{s}
\label{eff_Xsection_coann_real_S}
\end{align}
Here $g_S=1$ and $g_F=2$  denote the internal degrees of freedom of the scalar dark matter and the fermion mediator, respectively. The flavour multiplicity is treated explicitly through the factor $n_f=3$, which accounts for the three exactly degenerate mediator copies transforming as a triplet under U(3)$_e$. Let us briefly discuss how the $n_f$ factor enters the effective cross section. For the scalar pair annihilation, we write the cross section in Eq.~\eqref{SS_ann_tree_level} for a single lepton flavour, hence the factor of $n_f$ enters the inclusive cross section in Eq.~\eqref{eff_Xsection_coann_real_S}. For coannihilation processes, the factor $n_f=3$ simply counts the three identical channels involving the mediator flavours, namely $SF_e$, $SF_\mu$ and $SF_\tau$, together with the corresponding charge-conjugate processes. The same occurs for the  mediator pair annihilations, where, however, the different flavour combinations entering $ \sigma_{F \bar{F}} \vrel $ and $ \sigma_{F F} \vrel $ are slightly more intricate. More specifically, the process $F \bar{F} \to e \bar{e}$ has to be decomposed in (indices are not summed): (i) $F_a \bar{F}_a \to e_a \bar{e}_a$ (Yukawa and gauge contributing diagrams); (ii) $F_a \bar{F}_a \to e_b \bar{e}_b$ with $a \neq b$ (only $s$-channel gauge diagram), $F_a \bar{F}_b \to e_a \bar{e}_b$ with $a \neq b$ (only $t$-channel Yukawa diagram). The remaining processes of mediator-antimediator annihilations into gauge bosons, $Z h$, left-handed leptons and quarks bring in no complication and simply count with an overall factor of $n_f$ due to the flavour multiplicity.

Eq.~\eqref{eff_Xsection_coann_real_S} clearly illustrates the role of the compressed spectrum: both the coannihilation processes involving one dark matter particle and one mediator, as well as the pair annihilation of the mediators, are increasingly important for smaller $\Delta m$; otherwise, they become exponentially suppressed as the mass splitting increases.
For self-conjugate (real) scalar dark matter, mediator pair annihilations also receive contributions from the processes $FF\to ee$ and the corresponding charge-conjugate channel $\bar{F}\bar{F}\to\bar{e}\bar{e}$, collectively denoted by $\sigma_{FF}\vrel$ in Eq.~\eqref{eff_Xsection_coann_real_S}. These annihilation channels are absent in the case of complex scalar dark matter and represent one of the few qualitative differences between the two scenarios as far as the annihilations in the early universe are concerned. The expression for the effective cross section for $S \neq S^*$ is given in App.~\ref{app:cross_sections_relic}. 

Before discussing the coannihilation channels in some detail, it is instructive to recall the annihilation cross sections for dark matter pair annihilation. In the limit of vanishing lepton masses, the leading contributions are of order $\mathcal{O}(\vrel^4)$ for real scalar DM and $\mathcal{O}(\vrel^2)$ for complex scalar DM. This approximation is well justified in the present analysis, where the dark-sector masses are $m_S,m_F\geq 250$~GeV. Therefore, the $s$-wave velocity independent terms, which features an $(m_{e_a}/m_S)^2$ suppression,  are subleading to the velocity suppressed terms. The corresponding cross sections are
\begin{eqnarray}
\label{SS_ann_tree_level}
    \sigma_{SS}\vrel
    =
    \frac{y^4m_S^6}{60\pi(m_S^2+m_F^2)^4}\vrel^4,
    \qquad
    \sigma_{SS^\ast}\vrel
    =
    \frac{y^4m_S^2}{48\pi(m_S^2+m_F^2)^2}\vrel^2.
\end{eqnarray}
As anticipated, these cross sections are understood as for DM pairs annihialting into a single lepton flavour. Our result agrees with earlier determinations in Refs.~\cite{Bai:2014osa,Giacchino:2013bta,Kavanagh:2018xeh}, where the case with $n_f=1$ was considered. The strong velocity suppression of dark matter pair annihilation, particularly in the case of real scalar dark matter, considerably reduces its contribution to the effective annihilation cross section during freeze-out. As a consequence, coannihilation processes play a crucial role in reproducing the observed dark matter relic abundance. These processes are instead computed in our work. 

All coannihilation cross sections have been derived analytically and independently implemented using \texttt{FeynCalc} \cite{Shtabovenko:2020gxv}. The corresponding expressions are collected in App.~\ref{app:cross_sections_relic}. At tree level, the coannihilation process $SF$ scales as $g_1^2y^2$, whereas mediator pair annihilations exhibit a richer structure, with contributions proportional to $g_1^4$ (pure gauge), $g_1^2y^2$, and $y^4$ (pure Yukawa). Most importantly, all coannihilation channels possess a leading $s$-wave contribution and are therefore unsuppressed 
in the non-relativistic limit relevant for the freeze-out.

A further ingredient that must be accounted for concerns the annihilation of the charged fermion mediators. Slowly moving non-relativistic particles interacting through long-range forces mediated by vector gauge bosons experience important near-threshold effects, most notably the Sommerfeld enhancement~\cite{Hisano:2004ds,Iengo:2009ni,Feng:2010zp}, for an attractive potential, and the formation of bound states~\cite{Detmold:2014qqa,vonHarling:2014kha,Petraki:2015hla}. Both mechanisms increase the depletion rate of the mediator population via a modification of the effective annihilation cross section entering the Boltzmann equation. Although these effects are often neglected in phenomenological analyses of leptophilic dark matter, they are consistently included in our numerical treatment. Their impact is important in the present model, where the annihilation of scalar dark matter pairs is velocity suppressed and the relic abundance is largely controlled by coannihilation processes. As a consequence, the early-Universe dynamics of the fermion mediator play a crucial role throughout the compressed region of parameter space explored in this work. 

At temperatures below the electroweak crossover, the fermion mediator interacts with both the photon and the $Z$ boson. Consequently, both gauge bosons contribute to the long-range potential responsible for the Sommerfeld factors and bound-state formation. The contribution from the $Z$ boson is, however, strongly suppressed by its finite mass, which renders the corresponding Yukawa potential short ranged. The long-range dynamics are therefore dominated by photon exchange. It is worth stressing that, in the case of real scalar dark matter, the $F F$ and $\bar{F} \bar{F}$ annihilations experience instead a repulsive potential and, therefore, the Sommerfeld factor decrease the corresponding annihilation process. Moreover, bound states do not form in the repuslsive channel.

The distinction between the photon and the $Z$ boson ceases for sufficiently heavy dark matter. Since freeze-out occurs at a temperature $T_f\simeq m_S/25$, dark matter masses above approximately $4~\mathrm{TeV}$ correspond to freeze-out temperatures exceeding the electroweak crossover temperature, $T_{\rmi{c},\rmii{EW}}\simeq160~\mathrm{GeV}$. In this regime the electroweak symmetry is restored, and the mediator interacts with the hypercharge gauge boson $B_\mu$ rather than with the photon and the $Z$ boson. Consequently, the Sommerfeld enhancement and bound-state formation must be computed in the unbroken Standard Model.

As for the photon and $B_\mu$ exchange, we can use the analytic expression of the Sommerfeld factor, as well as for the bound-states contribution. We employ the modern treatment of non-relativistic effective field theories as presented in Ref. \cite{Biondini:2023zcz}, whereas the inclusion of excited bound states with transition follows the discussion in Ref.~\cite{Garny:2021qsr}. For the $Z$ boson, whose effects are milder due to the screening radius $\sim 1/m_Z$, we implement the numerical strategy for the extraction of the Yukawa Sommerfeld factor and bound-state effects as outlined in Refs. \cite{Petraki:2016cnz,Bollig:2021psb,Biondini:2023ksj}.

\subsection{Higher-order correction}
\label{sec:higher_order_S}

The SMEFT constraints can become increasingly relevant for large values of the Yukawa coupling $y$, which we take as large as the conventional upper limit  $y = \sqrt{4\pi}$ \cite{Cahill-Rowley:2015aea,Garny:2015wea}. A large Yukawa coupling also favours the sensitivity of direct and indirect detection to our model, with the nuclear recoil cross section and the annihilation into photons proportional to $y^4$. This motivates us to assess the impact of higher-order corrections in the Yukawa coupling on the annihilation processes entering the relic-density calculation in order to provide, at least from the cosmological side, a more accurate determination of the parameter space in the large-$y$ window.
\begin{table}[t!] \centering \renewcommand{\arraystretch}{1.35} \setlength{\tabcolsep}{12pt} \begin{tabular}{l c l} \hline Process & Order & Features (for $m_{e_a}=0$) \\ 
\hline 
$S S \to e_a \bar{e}_a$ & tree level & $d$-wave; $y^4$ \\ 
\hline $S S \to e_a \bar{e}_a \gamma$ & radiative $2\to3$ & $s$-wave; $y^4 \alpha_{\rmii{QED}}$ \\ $S S \to \gamma\gamma,\gamma Z$ & one loop & $s$-wave; $y^4 \alpha_{\rmii{QED}}^2$ \\ \hline $S S \to e_a \bar{e}_a $ & one loop & $d$-wave; $y^6$ and $y^4 \alpha_{\rmii{QED}}$ \\ 
\hline \end{tabular} \caption{Annihilation processes relevant to the discussion of higher-order effects. The radiative channels $S S\to e_a \bar{e}_a \gamma$, $S S\to\gamma\gamma$ and $S S\to\gamma Z$ were studied in Ref.~\cite{Ibarra:2014qma}, whereas the virtual Yukawa corrections to the $2\to2$ processes are computed in this work. We provide some comments for each process, including the leading wave and order of couplings.} \label{tab:annihilation_processes_NLO} \end{table}

Let us clarify early on what we focus on and the subclass of corrections we shall derive. We do not attempt a complete next-to-leading-order calculation of all $2 \to 2$ processes entering \eqref{eff_Xsection_coann_real_S}, the cross sections of which, at tree level, are listed in App.~\ref{app:cross_sections_relic}. Instead, we compute the virtual correction only to the dark matter annihilation process $S S \to e_a \bar{e}_a$.
More specifically, we will limit ourselves to considering virtual corrections to DM annihilation and can schematically organize our result as:
\begin{equation}
\sigma_{SS} \vrel = \sigma_{SS}\vrel \big\vert_{\rmi{tree}} + \Delta \sigma_{SS} \vrel \big\vert_{\rmii{NLO}} \, .
\label{eq:xs_NLO}
\end{equation}
In our NLO correction, the interference between the tree-level and one-loop amplitudes is considered. The squared one-loop amplitude is indeed formally higher order in the couplings. A complete one-loop improvement of all $2 \to 2$ processes is beyond the scope of the present work; a recent analysis in this direction for other DM models can be found in Ref.~\cite{Brica:2026nwi}.

For completeness, it is useful to distinguish the virtual corrections considered here from the radiative annihilation processes that have been studied previously for the same model \cite{Ibarra:2014qma,Giacchino:2013bta}. In the limit of vanishing lepton masses, the tree-level annihilation of real scalar dark matter, $S S \to e_a \bar{e}_a$
is $d$-wave suppressed, cf.~\eqref{SS_ann_tree_level}. This suppression can be lifted by radiative processes. In particular, Ref.~\cite{Ibarra:2014qma} showed that internal bremsstrahlung,
$S S \to e_a \bar{e}_a\gamma$, provides an $s$-wave contribution and can therefore be important for the relic-density calculation and, more prominently, for indirect detection in the compressed regime. The same work also considered the one-loop processes $S S \to \gamma\gamma$ and $S S \to \gamma Z$. These contributions correspond to additional annihilation channels and are therefore conceptually distinct from the virtual corrections to the $2\to2$ processes considered here.
The different contributions relevant for the present discussion are summarized in Tab.~\ref{tab:annihilation_processes_NLO}. 

We compute the one-loop corrections in dimensional regularization with
$d=4-2\epsilon$. Ultraviolet divergences are removed by introducing the appropriate counterterms for the fields, masses, and couplings. We employ an on-shell renormalization scheme for canonically normalized fields; this way, radiative corrections for the wave functions are not necessary. Masses and couplings are instead renormalized in the $\overline{\mathrm{MS}}$ scheme. Schematically, the bare model's bare Lagrangian is rewritten as the renormalized Lagrangian plus the counterterms:
\begin{equation}
\mathcal{L}_{\rmii{BSM}}
=
\mathcal{L}_{\rmii{SM}}^{(r)}
+
\mathcal{L}_{S}^{(r)}
+
\mathcal{L}_{F}^{(r)}
+
\mathcal{L}_{\rmi{portal}}^{(r)}
+
\delta\mathcal{L}_{\rmii{SM}}
+
\delta\mathcal{L}_{S}
+
\delta\mathcal{L}_{F}
+
\delta\mathcal{L}_{\rmi{portal}} , ,
\label{eq:lag_schem_NLO}
\end{equation}
where the superscript $(r)$ denotes quantities expressed in terms of renormalized fields and parameters, while the $\delta\mathcal{L}_i$ collect the corresponding counterterms. We have explicitly checked that no $SS\gamma$ and $SSZ$ dimension-four vertices are generated, and this is expected since such vertices are forbidden by gauge invariance. 

The calculation is performed through an automated pipeline based on
\texttt{FeynRules}~\cite{Alloul:2013bka},
\texttt{FeynArts}~\cite{Hahn:2000kx},
\texttt{FeynCalc}~\cite{Shtabovenko:2020gxv},
and \texttt{Package-X}~\cite{Patel:2015tea}.

\begin{figure}[t!]
\begin{minipage}{0.30\textwidth}
\centering
\includegraphics[width=\linewidth]{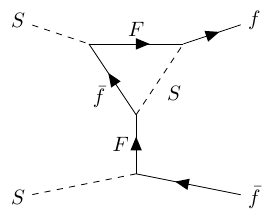}
\end{minipage}\hfill
\begin{minipage}{0.30\textwidth}
\centering
\includegraphics[width=\linewidth]{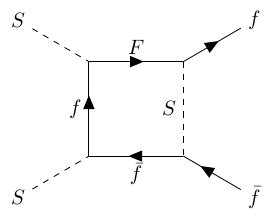}
\end{minipage}\hfill
\begin{minipage}{0.30\textwidth}
\centering
\includegraphics[width=\linewidth]{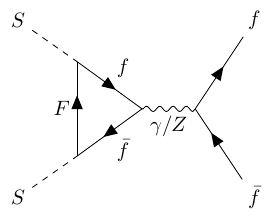}
\end{minipage}\hfill
\caption{Representative one-loop diagrams contributing to scalar dark matter annihilation into leptons.
}
\label{fig:1loop_SS}
\end{figure}

\subsubsection{Scalar dark matter annihilation}
\label{sec:higher_order_SS}

 As discussed above, the tree-level process $S S \to e_a \bar{e}_a$ is $d$-wave suppressed in the limit of vanishing lepton masses. In the same approximation, we find that all one-loop corrections retain the $d$-wave suppression, i.e. they do not lift the velocity suppression of the $2\to2$ process.
Their relevance is instead associated with the potentially large value of the Yukawa coupling.
A representative subset of the one-loop diagrams contributing to scalar dark matter annihilation is shown in Fig.~\ref{fig:1loop_SS}; for real scalar dark matter, the diagrams obtained by crossing the initial-state lines are included as well.

Our computation includes one-loop corrections mediated by the photon and the $Z$ boson. It is convenient to track the impact of these corrections through a dimensionless parameter $\tau_{\scriptscriptstyle Z}$, defined as
\begin{equation}
\tau_{\scriptscriptstyle Z} = \dfrac{m_Z^2}{m_S^2} \, .
\end{equation}
When expressed in terms of $\rho=m_S^2/m_F^2$ and $\tau_{\scriptscriptstyle Z}$, the ratio of the radiative to the tree-level cross section becomes independent of the absolute dark matter mass scale, in full analogy with the bremsstrahlung contribution, which depends on $\rho$ alone \cite{Ibarra:2014qma}.

\begin{figure}[t!]
\centering
\centering
\includegraphics[width=0.65\textwidth]{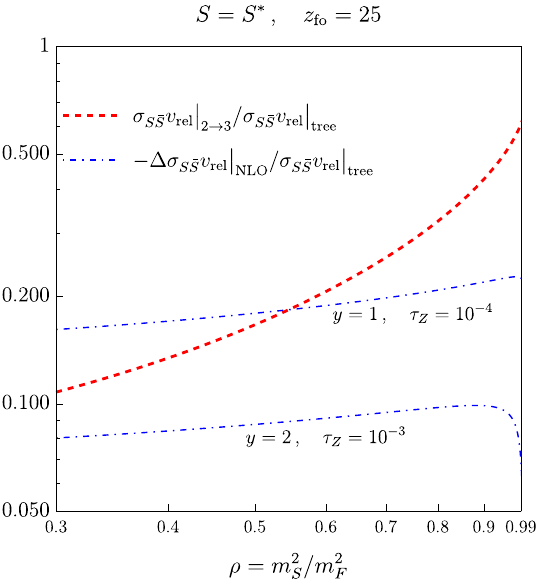}
\caption{The red curve shows the gamma-ray contribution computed, e.g., in \cite{Ibarra:2014qma}. The two blue curves show the one-loop electroweak correction computed in this work, for two representative values of $\tau_Z$. 
}
\label{fig:one_loop_DM_DM_and_Ibarra}
\end{figure}

It is natural to how our result compare with the radiative $2 \to 3$ process \cite{Ibarra:2014qma}. To address this matter, in Fig.~\ref{fig:one_loop_DM_DM_and_Ibarra} we compare the different radiative contributions, each shown as the absolute value of its ratio to the tree-level thermally averaged cross section \eqref{SS_ann_tree_level}.
Although the bremsstrahlung contribution dominates for $\rho$ close to unity, the one-loop correction to the $2 \to 2$ process becomes increasingly important as $\rho$ decreases. Therefore, their effects are complementary. This behaviour is also manifest at the level of the analytic expression, obtained in the limit $\rho \to 0$, $\tau_{\scriptscriptstyle Z} \to 0$:
\begin{equation} \begin{split} \Delta & \sigma \vrel^{\rm NLO} \; \simeq \; -\frac{7\,\vrel^4\, y^6\, \rho^4}{3840\, m_S^2\, \pi^3} \\[8pt] &+\; \frac{\vrel^4\, y^4\, \alpha_{\rmii{QED}}}{m_S^2\, \tau_{\scriptscriptstyle Z}^2 \rho^{12}} \left[ C_1 \left(1 - 11 \tau_{\scriptscriptstyle Z} + 66\tau_{\scriptscriptstyle Z}^2\right) + C_2 \left(-1+2\sw^2\tau_{\scriptscriptstyle Z} \right) \tau_{\scriptscriptstyle Z}^2 \rho^{14} \left(\pi^2 - 6\, \mathrm{Li}_2\!\left(\tfrac{1}{1+\rho}\right)\right) \right]\,, \end{split} \label{eq:compact_NLO} \end{equation}
  where $\text{Li}_2(x)$ is the dilogarithm function, $C_{1,2}$ are combinations of the Weinberg angle and numerical coefficients, and read
  \begin{equation} C_1 = \frac{-3\left(-1-8\sw^2+4\sw^4\right)}{47185920\,\pi^2\,\sw^2\left(-1+\sw^2\right)}\, , \qquad C_2 = \frac{16384\left(-1+2\sw^2\right)}{47185920\,\pi^2\,\sw^2\left(-1+\sw^2\right)}\, . \end{equation}
  Eq.~\eqref{eq:compact_NLO} shows that the electromagnetic correction dominates for small $\rho$. Notably, the entire $\tau_{\scriptscriptstyle Z}$ dependence resides in this electromagnetic term, which drives $\Delta \sigma \vrel \vert_{\rmii{NLO}}$ to large values as $\tau_{\scriptscriptstyle Z}$ decreases. We have evaluated the thermally averaged cross section, both at tree level and including the one-loop correction, for a fixed freeze-out benchmark temperature of $z_{\rmi{fo}} = 25$. We also note that the NLO correction to the $2 \to 2$ DM annihilation cross section is always negative. 

For completeness, we stress that the cross section defined in \eqref{eq:xs_NLO} is not an infrared-safe quantity: divergences appear in the $\tau_{\scriptscriptstyle Z} \to 0$ limit. These would be regulated by retaining a nonzero lepton mass $m_e$, i.e.\ the resulting logarithms would depend on $\tau_e \equiv m_e^2/m_S^2$. We refrain from performing the complete one-loop computation with this infrared regulator, since both the results of \cite{Ibarra:2014qma} and our phenomenological analysis are carried out in the chiral limit\footnote{For the sake of completeness we stress this quantity  can be made IR finite once we are inclusive enough in the final state for the KLN theorem. This means and IR safe quantity for $SS\to l \bar{l}$ would require to sum the IR poles coming from real emission diagrams $SS \to l \bar{l}\gamma$, moreover these contribution should be of the order $\vrel^4$.}. In any case, the limit for $\tau_{\scriptscriptstyle Z} \to 0$ is not reached, and for $\mathcal{O}(1)$ TeV dark matter, $\tau_{\scriptscriptstyle Z}$ remains reasonably larger than zero.

Given that the NLO corrections retain the velocity suppression, we include them only in the estimation of the relic density and not in the indirect-detection phenomenology.
\subsection{Modified cosmological history}
\label{sec:mod_cosm}

The relic-density calculation usually assumes the standard
radiation-dominated cosmological history during the DM freeze-out. This assumption is well
motivated at temperatures of order a few MeV, where Big Bang
Nucleosynthesis (BBN) constrains the expansion rate of the Universe.
At higher temperatures, however, the cosmological history is less
directly constrained, and the expansion rate during dark matter
freeze-out can differ from the standard radiation-dominated
expectation~\cite{DEramo:2017gpl,Biondini:2023hek}.

Following Ref.~\cite{DEramo:2017gpl}, we parameterise the modified
cosmological history by introducing an additional component $\varphi$
whose energy density redshifts as
\begin{equation}
    \rho_\varphi \propto a^{-(4+n)}\,,
\end{equation}
where $a$ is the scale factor and $n>0$. The total energy density is
then given by $ \rho_{\rm tot}(T)
    =
    \rho_{\rm rad}(T)+\rho_\varphi(T)$ and we define the reference temperature $T_{\rm ref}$ through
\begin{equation}
    \rho_\varphi(T_{\rm ref})
    =
    \rho_{\rm rad}(T_{\rm ref})\,.
\end{equation}
For temperatures above $T_{\rm ref}$, the $\varphi$ component dominates
the expansion rate, leading to a Hubble rate larger than in the
standard radiation-dominated cosmology. The modification becomes more
pronounced for smaller values of $T_{\rm ref}$~\cite{DEramo:2017gpl}.
The modified expansion history therefore enters our relic-density
calculation solely through the Hubble rate $H$ in Eq.~\eqref{BE_start},
while the particle-physics ingredients entering
$\langle\sigma_{\rm eff}v_{\rm rel}\rangle$ remain unchanged.
The explicit form of the effective degrees of freedom can be written
in terms of a generalized number of relativistic degrees of freedom
for the modified cosmologies $(T_{\rm ref},n)$,
\begin{equation}
    g_{\rm eff}^{\varphi}(T,T_{\rm ref},n)
    \equiv
    g_{\rm eff}(T)
    +
    g_{\rm eff}(T_{\rm ref})
    \left(
        \frac{h_{\rm eff}(T)}
             {h_{\rm eff}(T_{\rm ref})}
    \right)^{\frac{4+n}{3}}
    \left(
        \frac{T}{T_{\rm ref}}
    \right)^n \, ,
\end{equation}
which is then inserted into the Hubble rate. Here, $g_{\rm eff}$ and
$h_{\rm eff}$ are, respectively, the energy and entropy relativistic
degrees of freedom.

For illustration, we consider the case $n=2$, corresponding to a
kination regime that can arise in quintessence
scenarios~\cite{Caldwell:1997ii,Sahni:1999gb}. We take
$T_{\rm ref}=1~\mathrm{GeV}$ as a representative benchmark. This value
lies above the QCD crossover and below the freeze-out temperatures
relevant for the dark matter masses considered in this work, such that
the modified expansion rate is active during freeze-out while avoiding
a parametrisation deep in the non-perturbative QCD regime.
Smaller values of $T_{\rm ref}$ are also possible, but are constrained
by BBN. For the parametrisation adopted here, the corresponding
constraint can be expressed approximately as $T_{\rm ref} \gtrsim (15.4)^{1/n}\,\mathrm{MeV}$, as obtained in Ref.~\cite{DEramo:2017gpl}. We therefore restrict our
illustrative analysis to cosmological histories that satisfy this
condition.
We discuss the impact on the DM relic density in
Sec.~\ref{sec:results}.

\section{Experimental searches}
\label{sec:constraints}

In this section we address the experimental approaches to particle dark matter detection. As we consider dark matter masses $m_S > 250$ GeV, the most severe constraints are imposed by direct detection, which is the next subject of the paper in Sec.~\ref{sec:DD}, and then we include a briefer discussion on collider searches and indirect searches in Sec.~\ref{sec:coll_and_ID}. 
\subsection{Direct Detection}
\label{sec:DD}

\begin{figure}[t!]
    \centering
    \begin{minipage}{0.3\textwidth}
        \centering
        \includegraphics[width=\textwidth]{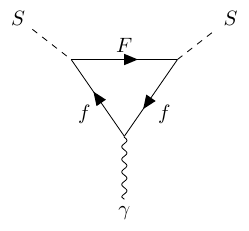}

    \end{minipage}%
    \hspace{0.15\textwidth}
    \begin{minipage}{0.3\textwidth}
        \centering
        \includegraphics[width=\textwidth]{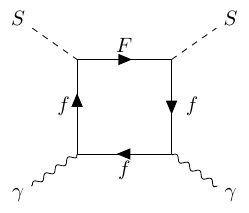}

    \end{minipage}
    \caption{Representative one-loop diagrams contributing to the matching onto the effective dark matter--photon operators. The triangle topology (left) contributes to the charge-radius coefficient $c_\gamma$, whereas the box topology (right) generates the Rayleigh coefficient $c_{\gamma\gamma}$. 
    }
    \label{fig:one_loop_DD_example}
\end{figure}

Since the dark matter particle is electrically neutral and interacts at tree level only with charged leptons through the exchange of the vector-like fermion mediator, elastic scattering off nuclei is absent at tree level. The leading contribution therefore arises at one loop through diagrams involving the charged mediator and the SM lepton, see Fig.~\ref{fig:one_loop_DD_example}, which induce effective interactions between dark matter and the electromagnetic field. This mechanism is analogous to that encountered in fermionic dark matter and lepton-portal models \cite{Kopp:2014tsa,Ibarra:2015fqa}, although the effective operators are different  due to the scalar nature of the dark matter candidate.

We note that the lepton-portal interaction also induces tree-level scattering of dark matter off electrons, $S e\to S e$, through $t$-channel exchange of the mediator $F$. While this provides an additional direct-detection channel, its treatment requires a dedicated calculation of the electron-ionisation rate in the detector material \cite{Catena:2019gfa}, including the corresponding atomic response functions; we therefore leave the analysis of electron recoils for future work and understand the constraints from direct detection as conservative. 

At the momentum transfers relevant for underground direct-detection experiments,
$q\lesssim \mathcal{O}(100)\,\mathrm{MeV}$, the full one-loop amplitude can be expanded in powers of the small momentum transfer and matched onto a low-energy effective theory. The latter is organised in inverse powers of the dark matter mass, for which it holds $q^2 /m_S^2 \ll 1$. In our work, the matching is performed by keeping the full dependence on the internal masses.  For a complex scalar dark matter particle, gauge invariance allows three independent electromagnetic operators at dimension six \cite{Bai:2014osa,Brod:2017bsw,Kavanagh:2018xeh}
\begin{align}
\mathcal{L}_{S\neq S^*}^{(6)}
=
\frac{c_\gamma}{m_S^2}\,
i\,S^*
\overleftrightarrow{\partial_\mu}
S\,\partial_\nu F^{\mu\nu}
+
\frac{2 c_{\gamma\gamma}}{m_S^2}\,
S^*S\,F_{\mu\nu}F^{\mu\nu} \, +  \dfrac{d_{\gamma \gamma}}{m_S^2} S^2\epsilon_{\mu\nu\rho \sigma}F^{\rho \sigma}F^{\mu\nu}.
\label{eq:LeffDD_complex}
\end{align}
The matching coefficient $c_\gamma$ parametrizes the electromagnetic charge-radius interaction, while $c_{\gamma\gamma}$ and $d_{\gamma \gamma}$ corresponds to the scalar Rayleigh interactions, the latter involving the dual of the photon field. Interactions of dark matter with two photons, commonly referred to as Rayleigh interactions, were first studied in the context of Majorana fermionic WIMPs \cite{Weiner:2012cb} and indirect detection \cite{Frandsen:2012db}. The charge-radius operator is a distinctive feature of complex scalar dark matter, since the current $i\,S^\dagger
\overleftrightarrow{\partial_\mu}
S$
vanishes identically for a self-conjugate (real) scalar field. Consequently, for real scalar dark matter \emph{only} the Rayleigh operators are present at order $1/m_S^2$ \cite{Bai:2014osa,Brod:2017bsw,Kavanagh:2018xeh}
\begin{align}
\mathcal{L}_{S =  S^*}^{(6)}
=
\frac{c_{\gamma\gamma}}{m_S^2}\,
S^2\,F_{\mu\nu}F^{\mu\nu} \, +  \dfrac{d_{\gamma \gamma}}{2\, m_S^2} S^2\epsilon_{\mu\nu\rho \sigma}F^{\rho \sigma}F^{\mu\nu}.
\label{eq:LeffDD_real}
\end{align}
In order to write a single expression for $c_{\gamma \gamma}$ and $d_{\gamma \gamma}$, we write an explicit factor of 2 accounting for the multiplicity of the diagrams in the complex dark matter case in Eq.~\eqref{eq:LeffDD_complex}. 

The charge-radius operator mediates the interaction through single-photon exchange with the nuclear electromagnetic current and generates the leading spin-independent scattering cross section. In contrast, the Rayleigh operators couple the dark matter simultaneously to two photons and contribute through two-photon exchange with the nucleus. Exemplary one-loop diagrams for the matching onto the effective operators are shown in Fig.~\ref{fig:one_loop_DD_example}.

Although both operators are induced at the same loop order, the Rayleigh contributions are expected to be numerically subleading because of the additional suppression in $\alpha_{\rmi{em}}$.  Consequently, the direct-detection phenomenology is largely controlled by the coefficient $c_\gamma$ in the case of complex DM, while the Rayleigh interaction provides only a small correction. The situation changes in the real scalar case, where the Rayleigh operator induces the leading contribution to the recoil cross section. 

In our work we determine $c_\gamma$, $c_{\gamma\gamma}$ and $d_{\gamma \gamma}$ by matching the relevant one-loop amplitudes of the DM leptophilic model \eqref{eq:lag_schem} onto the effective operators \eqref{eq:LeffDD_complex} and \eqref{eq:LeffDD_real}. Improving on existing results in the literature, we retain the full dependence on the mediator and lepton masses, encoded in the two dimensionless ratios of squared masses
\begin{align}
    \rho = \frac{m_S^2}{m_F^2} \, , \quad \tilde{\rho}= \frac{m_e^2}{m_F^2} \, .
\end{align}
We then provide more general results for the direct detection, including the kinematic region of a moderately degenerate and compressed spectrum $\Delta m/m_S <1$, going beyond the limit $m_F \gg m_S$. This is also the regime that is favoured by dark matter cosmological abundance, namely coannihilation assisted, and relevant for the interplay with the constraints from global SMEFT fits. 

The extraction of the matching coefficients depends, however, on the lepton flavour. As discussed in Ref.~\cite{Kopp:2014tsa}, the distinction originates from the relative size of the typical momentum transfer and the mass of the SM lepton running in the loop. For the muon and tau lepton, the relevant hierarchy of scales is\footnote{For nuclear recoils the momentum transfer is related to the recoil energy through \(|\bm{q}|^2=2m_{N} E_R\). The recoil energies relevant for current xenon-based experiments correspond to momentum transfers of order tens of MeV, with \(\sqrt{-q^2} \simeq |\bm{q}|\simeq50\) MeV providing a representative benchmark. The hierarchy for the muon case may not always be realized for large nuclear recoil energies.}
\begin{equation}
    q^2 \ll m_{\mu,\tau}^2 \ll m_S^2 \lesssim m_F^2 \, .
    \label{eq:hierachy_heavy_leptons}
\end{equation}
The small-momentum-transfer limit can therefore be taken safely when expanding the amplitude, allowing us to neglect terms suppressed by powers of $q^2/m_{\mu,\tau}^2$. For the electron case, on the other hand, we follow the strategy of Ref.~\cite{Kopp:2014tsa} and neglect the electron mass, since the corresponding hierarchy of scales is 
\begin{equation}
     m_{e}^2 \ll q^2  \ll m_S^2 \lesssim m_F^2 \, .
\end{equation}
In this case, the leading dependence on the momentum transfer is retained, giving rise to logarithmic terms involving the ratio of the momentum transfer to dark matter mass.

The resulting matching coefficients read, respectively for the charged radius and Rayleigh operators, as follows 
\begin{align}
    c^{(\mu,\tau)}_\gamma  &= \dfrac{y^2 e \,  Q_F}{24 \pi^2}\, \,\mathcal{F}_{\gamma}^{(\mu,\tau)}(\rho ,\tilde{\rho}) \, , \quad c^{(\mu,\tau)}_{\gamma\gamma}\, = \,  \dfrac{y^2 \, e^2 Q_F^2\,\mathcal{F}^{(\mu,\tau)}_{\gamma \gamma}(\rho ,\tilde{\rho})}{24 \pi^2} \, , \quad d^{(a)}_{\gamma \gamma} =0 \, ,
    \label{eq:heavy_cDD}
    \\
   c^{(e)}_\gamma & = \dfrac{y^2 e \,  Q_F}{24 \pi^2}\, \,\mathcal{F}_{\gamma}^{(e)}(\rho ,\xi) \, , \quad~~~~~c^{(e)}_{\gamma\gamma}\, = \,  \dfrac{y^2 \, e^2 Q_F^2\,\mathcal{F}^{(e)}_{\gamma \gamma}(\rho ,\xi)}{24 \pi^2} \, ,
   \label{eq:light_cDD}
\end{align}
where $Q_F=-1$ is the electric charge of the fermion mediator, in units of $e$, equal to that of the SM charged leptons, whereas $\mathcal{F}^{(\mu,\tau)}_{\gamma}(\rho ,\tilde{\rho})$, $\mathcal{F}^{(\mu,\tau)}_{\gamma \gamma}(\rho ,\tilde{\rho})$, $\mathcal{F}^{(e)}_{\gamma}(\rho ,\xi)$ and $\mathcal{F}^{(e)}_{\gamma \gamma}(\rho,\xi)$ encode the resulting loop functions arising from the triangle- and box-diagram topologies respectively. Their explicit expression are given in the App.~\ref{app:DD}. In the electron case, there is no dependence on $\tilde{\rho}$, since we
have taken the limit $m_e\to0$. Instead, the result depends on the
transferred momentum through $\xi\equiv -q^2/m_F^2 >0$. In this limit, the
finite momentum transfer regulates the infrared logarithm in the amplitude
and the corresponding matching coefficient, similarly to the treatment
adopted in Ref.~\cite{Kopp:2014tsa}.\footnote{Strictly speaking, the logarithmic dependence on the momentum transfer is not expected in a conventional local matching coefficient, which should encode short-distance physics after infrared contributions common to the full and effective theories have been subtracted. Here, taking the limit \(m_e\to0\) leaves the momentum transfer as an infrared regulator, so that the resulting coefficient retains sensitivity to a scale below the mediator mass. It should therefore be understood as a momentum-dependent quantity obtained in this limit, rather than as a conventional Wilson coefficient defined by expanding in all light scales and setting them to zero. This residual infrared sensitivity should be kept in mind when interpreting the result.} 
For $m_F\gg m_S$, the loop functions in Eqs.~\eqref{eq:heavy_cDD} and \eqref{eq:light_cDD} reproduces
the corresponding large mediator--dark matter mass-splitting limits, the structural form of which is
in agreement with earlier results in the literature \cite{Kavanagh:2018xeh}. Moreover, we find agreement with Ref.~\cite{Kavanagh:2018xeh} for the absence of the matching coefficient of the CP-odd operator $d_{\gamma \gamma}^{(a)}$.\footnote{For the derivation of $c_\gamma$, we adopt the naive dimensional regularization (NDR) scheme. In this case, the calculation is unambiguous since the relevant traces contain at most four $\gamma$ matrices. The box diagrams contributing to $c_{\gamma\gamma}$ and $d_{\gamma\gamma}$ require more care, as their evaluation involves the treatment of $\gamma^5$ in dimensional regularization. Since $c_{\gamma\gamma}$ refers to a CP-even operator we can consistently choose a NDR scheme and no ambiguities arise there. We have explicitly checked that the coefficient of the CP-odd operator involving the dual photon field-strength tensor, $F_{\mu\nu}\widetilde{F}^{\mu\nu}$, vanishes both in NDR and in consistent schemes for the treatment of $\gamma^5$ and fermion-loop diagrams i.e. Breitenlohner-Maison-'t Hooft-Veltman (BMHV) scheme.}  
The matching has been obtained both via analytical derivations and automated tools, more specifically \texttt{FeynCalc} \cite{Shtabovenko:2020gxv} and \texttt{Package-X}~\cite{Patel:2015tea}.

The next step is to determine the dark matter--nucleus scattering cross section and extract the corresponding spin-independent dark matter--nucleon cross section. In the numerical analysis, we compare the latter with the mass-dependent upper limits reported by the LZ collaboration \cite{LZ:2024zvo}, which currently provide the most stringent direct-detection constraints. The quoted experimental limits assume isospin-conserving couplings, while our electromagnetic interaction is proton-only. We therefore use the published limits for comparison, with the resulting bounds understood as approximate; a dedicated recast accounting for the xenon proton-only response would be required for a precise exclusion. 

The treatment of the nuclear interaction, however, differs qualitatively for the two effective operators.
For the charge-radius operator, the interaction proceeds through the exchange of a single photon, which couples coherently to the nuclear electromagnetic current. The resulting nuclear recoil cross section, derived in App.~\ref{app:DD}, follows the standard formalism of Ref.~\cite{Ibarra:2015fqa}, where the finite-size effects of the nucleus are encoded in the spin-independent Helm nuclear form factor.

The Rayleigh operator couples the dark matter simultaneously to two
photons, giving rise to a coherent interaction with the nuclear
electromagnetic field. In contrast to the single-photon case, the
corresponding nuclear matrix element contains a genuine two-body
contribution involving pairs of protons. Coherent two-photon exchange
with the nucleus was studied in the context of Rayleigh dark matter in
Refs.~\cite{Weiner:2012cb,Frandsen:2012db}, where the nuclear response
was described by a phenomenological coherence form factor. The
two-proton contribution was subsequently formulated within an
effective field theory framework in Ref.~\cite{Ovanesyan:2014fha}, in
terms of a two-proton nuclear density and its associated form factor,
including proton--proton correlations. In the limit in which these
correlations are neglected, the latter formalism reduces to the
phenomenological result of Ref.~\cite{Weiner:2012cb}, up to the
corresponding convention for the nuclear coherence scale $Q_0$. It is
worth stressing that the two-proton correlation form factor is not
known with sufficient accuracy, and Ref.~\cite{Ovanesyan:2014fha}
finds that it can induce an $\mathcal{O}(1)$ correction to the nuclear
matrix element. This translates into a potentially significant
theoretical uncertainty in the Rayleigh contribution to the
direct-detection rate. In the present model, this uncertainty has a
negligible impact on the predicted direct-detection rate for complex
dark matter, since the charge-radius interaction provides the dominant
contribution to the scattering cross section. For real scalar dark
matter, however, the Rayleigh interaction constitutes the leading
electromagnetic contribution to direct detection.\footnote{Minor discrepancies are found and can be traced back to a
different definition of the nuclear coherence scale $Q_0$. We also
warn the reader about a misprint in Ref.~\cite{Weiner:2012cb}, where
the nucleon cross section is written with a factor $1/A^4$, instead of
$1/A^2$, cf.~\eqref{cross_SI_real}; the same typo is reproduced in
Ref.~\cite{Kavanagh:2018xeh}.}

The DM--nucleon cross sections that we compare with the experimental
spin-independent exclusion limits from the LZ collaboration are
\begin{align}
    &\sigma^{S \neq S^*}_{\rmii{SI}}
    =
    \alpha_{\rmii{QED}}
    \left(
    \frac{\mu^N_{\rmi{red}}}{m_S}
    \right)^2
    \frac{4 c_\gamma^2}{m_S^2}
    +
     \frac{8 Z^4}{A^2}
    \left(
    \frac{\alpha_{\rmii{QED}}}{\pi}
    \right)^2
    \left(
    \frac{\mu^N_{\rmi{red}} Q_0}{m_S^2}
    \right)^2
    \frac{c_{\gamma\gamma}^2}{m_S^2} \, ,
    \label{cross_SI_complex}
    \\
    &\sigma^{S = S^*}_{\rmii{SI}}
    =
    \frac{8 Z^4}{A^2}
    \left(
    \frac{\alpha_{\rmii{QED}}}{\pi}
    \right)^2
    \left(
    \frac{\mu^N_{\rmi{red}} Q_0}{m_S^2}
    \right)^2
    \frac{c_{\gamma\gamma}^2}{m_S^2} \, .
    \label{cross_SI_real}
\end{align}
Here $Q_0 = 0.5 (0.3+0.9 A^{1/3})^{-1}\,{\rm GeV}$ \cite{Ovanesyan:2014fha}, with $A=131$ and $Z=54$ denoting the
average mass number and atomic number of xenon. 
In accordance with
our MVF assumption, we have summed over the lepton flavours, such that
$c_\gamma = \sum_a c_\gamma^{(a)}$ and, analogously,
$c_{\gamma\gamma} = \sum_a c_{\gamma\gamma}^{(a)}$. If instead one considers
dark matter interacting with a single lepton flavour at a time, only the
corresponding matching coefficient should be used in
Eqs.~\eqref{cross_SI_complex} and \eqref{cross_SI_real}.

\subsection{Collider searches and indirect detection}
\label{sec:coll_and_ID}
Dedicated LHC studies of leptophilic $t$-channel portals remain scarce, and a recent review on $t$-channel models and their collider signatures focusses only a single benchmark~\cite{Arina:2025zpi}, namely Majorana dark matter coupled exclusively to right-handed muons. In this scenario, the mediators are produced through Drell--Yan processes and decay promptly throughout the thermally favoured region, giving rise to a final state with two opposite-sign muons and missing transverse energy. For small mass splittings, however, the leptons become soft, requiring either dedicated soft-lepton strategies or an ISR-boosted search. Recasting existing ATLAS analyses, the compressed-spectrum slepton search~\cite{ATLAS:2019lff} excludes $m_{\rmii{DM}} \lesssim 200$~GeV (for $m_{\rmi{med}} \simeq 2 m_{\rmii{DM}}$), while the ISR soft-lepton search reaches $m_{\rmii{DM}}\lesssim 150$~GeV at about $10\%$ mass splitting. These searches improve upon the LEP2 bounds, but leave a substantial part of the parameter space, in particular at higher masses and for intermediate splittings ($\delta m \approx 0.1$--$2$), unconstrained.
Since our analysis assumes $m_S \geq 250$GeV no constraint from collider searches enter the following comprehensive plots in Sec.~\ref{sec:results}.

The limited collider sensitivity to compressed leptophilic spectra motivates considering complementary probes. In particular, indirect-detection searches can remain sensitive to the same region of parameter space that is favoured by coannihilation, while relying on the present-day annihilation properties of the dark matter rather than on the direct production of the mediator at colliders~\cite{Olgoso:2025jot}. The relevance of these searches therefore depends on whether the annihilation channels that are not velocity suppressed provide an appreciable present-day signal.

Turning to indirect detection, searches for distinctive features in the gamma-ray spectrum provide a complementary probe of dark matter annihilation against astrophysical backgrounds~\cite{Bergstrom:1988fp,Rudaz:1986db,Srednicki:1985sf}. In the leptophilic model considered here, photons are produced through two processes: the $2\to3$ annihilation channel $SS\to e_a\bar e_a\gamma$, corresponding to internal bremsstrahlung, and the loop-induced $2\to2$ channels $SS\to\gamma\gamma$ and $SS\to Z\gamma$. The corresponding cross sections have been computed in Ref.~\cite{Ibarra:2014qma}, and we adopt their results, with the appropriate modification to account for the lepton-flavour multiplicity in our model.

In contrast to the velocity-suppressed tree-level annihilation discussed in Eq.~\eqref{SS_ann_tree_level}, these radiative processes contain velocity-independent contributions and can therefore remain relevant in the present-day Universe. Although they are suppressed by powers of $\alpha_{\rmii{QED}}$ relative to the tree-level annihilation, this suppression can be partially compensated by enhancements associated with small mass splittings between the dark matter particle and its mediator. Consequently, radiative annihilation channels can provide relevant indirect-detection probes of regions where the tree-level annihilation rate is strongly velocity suppressed.

We evaluate the present-day annihilation rates along the cosmologically viable parameter space. Since the model is described by three independent parameters, $m_S$, $\delta m$, and $y$, we fix one parameter at a time and determine the corresponding curves in the remaining two-dimensional parameter plane that reproduce the observed dark matter relic abundance, $\Omega_{\rmii{DM}}h^2=0.120(12)$~\cite{Planck:2018vyg}. The resulting relic-density contours define the parameter space on which we evaluate the radiative annihilation cross sections. For the velocity-independent contributions considered here, we use velocity-independent cross sections from Ref.~\cite{Ibarra:2014qma}, which provide the appropriate present-day rates.

We compare these predictions with gamma-ray constraints from Fermi-LAT~\cite{Fermi-LAT:2015kyq} and H.E.S.S.~\cite{HESS:2018cbt,HESS:2026ila}, distinguishing between the two experiments and the corresponding spectral signatures. In the mass range $250~{\rm GeV}\lesssim m_S\lesssim294~{\rm GeV}$, the Fermi-LAT limits provide coverage of only a restricted portion of the parameter space considered in this work. In this interval, we compare the sum of the radiative annihilation cross sections for the processes $SS\to e_a\bar e_a\gamma$, $SS\to\gamma\gamma$ and $SS\to Z\gamma$
with the corresponding Fermi-LAT limits. Since the three channels give rise to different photon spectra, this comparison does not constitute a dedicated spectral analysis or a statistically consistent combined exclusion. Rather, it provides an indicative estimate of the sensitivity of Fermi-LAT to the total radiative annihilation rate in this limited mass range. 

For $m_S\gtrsim310~{\rm GeV}$, we instead use the latest H.E.S.S. limits from searches for monochromatic gamma-ray lines, which extend into the multi-TeV regime, and use the most recent data release \cite{HESS:2026ila}. The relevant signal arises from the one- and two line-producing channels, $SS\to\gamma\gamma$ and $SS\to Z\gamma$. We compare the predicted line cross section,
\begin{align}
\langle \sigma \vrel \rangle_{\rm line}
=\langle\sigma \vrel \rangle_{\gamma\gamma}
+\frac{1}{2}\langle\sigma \vrel \rangle_{Z\gamma},
\end{align}
with the corresponding H.E.S.S. upper limit at the dark matter mass $m_S$.  The factor of $1/2$ accounts for the different photon multiplicities in the two final states. Unlike the indicative Fermi-LAT comparison, this provides a direct comparison with the observable constrained by the H.E.S.S. line search. 

The two experimental analyses considered here do not provide coverage in the intermediate mass range $294~{\rm GeV}\lesssim m_S\lesssim310~{\rm GeV}$. We therefore do not derive indirect-detection constraints from these searches in this interval.

\section{Combined phenomenology and parameter space}
\label{sec:results}
In this section, we combine the cosmologically viable parameter space of the model with the constraints from direct and indirect detection, as well as those from global SMEFT fits.

\begin{figure}[t!]
    \centering
    \includegraphics[width=0.48\textwidth]{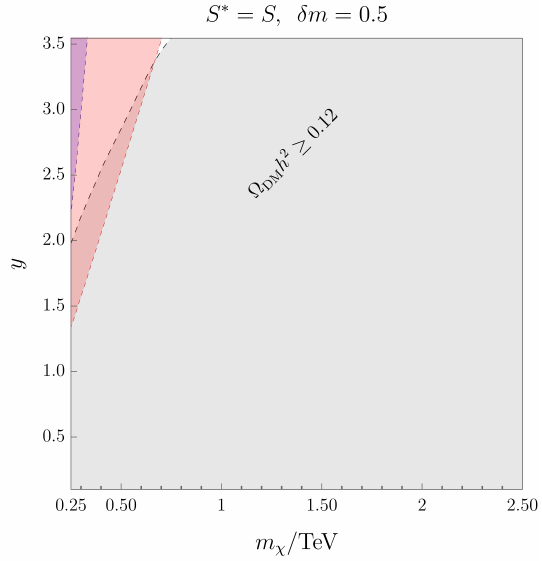}
    \hfill
    \includegraphics[width=0.48\textwidth]{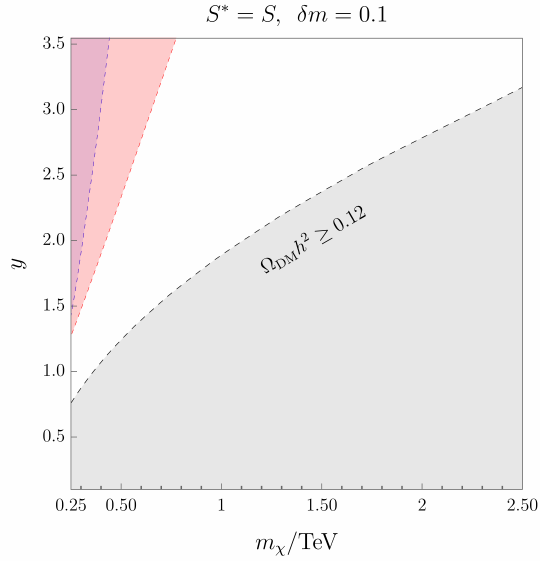}

    \vspace{0.5cm}

    \includegraphics[width=0.48\textwidth]{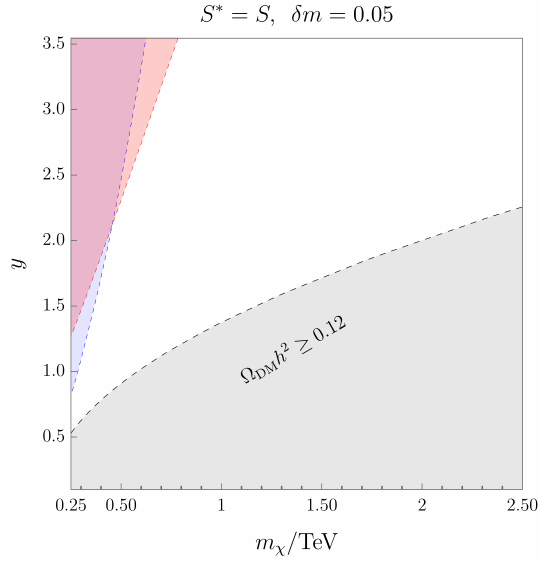}
    \hfill
    \includegraphics[width=0.48\textwidth]{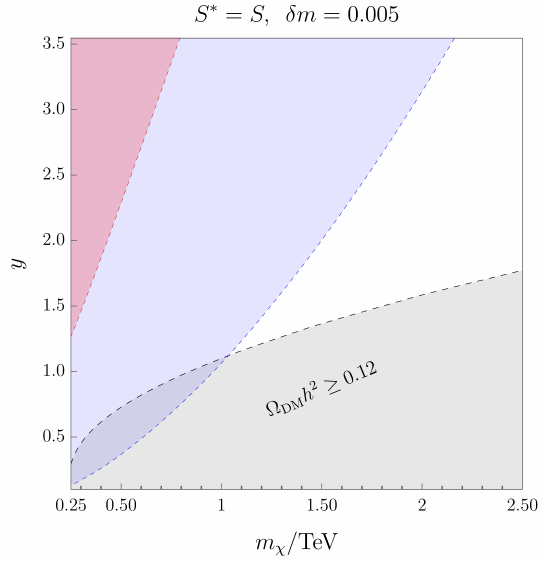}
    \caption{The parameter space for four representative values of the relative mass splitting. The gray-shaded regions are excluded by the observed relic abundance, while the red and light-blue regions are excluded by global SMEFT fits and direct-detection experiments, respectively.
}
    \label{fig:4panels_y_ms}
\end{figure}

We begin by combining the main results of our analysis: the SMEFT matching coefficients, the relic density, and the direct-detection predictions, computed for general mass splittings. These are presented in Secs.~\ref{sec:SMEFT_and_DM}, \ref{sec:relic} and~\ref{sec:DD}, respectively. We explore their complementarity in the \((m_S,y)\) plane for different values of the mass splitting. Fig.~\ref{fig:4panels_y_ms} shows the results for real scalar dark matter, considering four representative values of the relative mass splitting, $\delta m=0.5, \,0.1,\,0.05$ and \(0.005\). The complex scalar scenario is entirely excluded by direct-detection constraints, owing to the sizable contributions from the charge-radius operators; see Eq.~\eqref{eq:LeffDD_complex}. We therefore do not consider it in the summary plots below.

\begin{figure}[t]
\centering
\includegraphics[width=0.495\linewidth,height=0.57\linewidth]{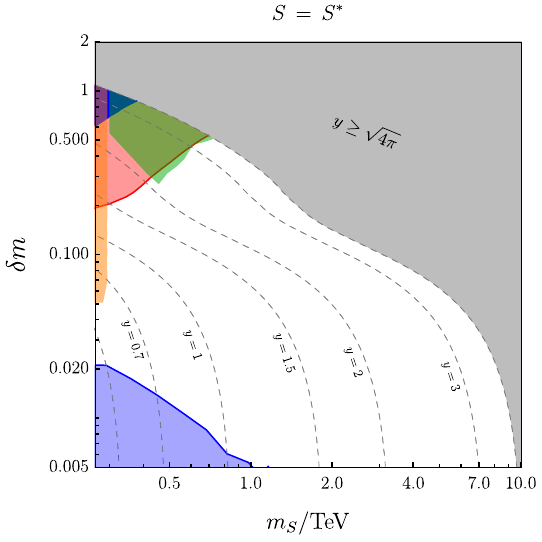}
\includegraphics[width=0.495\linewidth,height=0.57\linewidth]{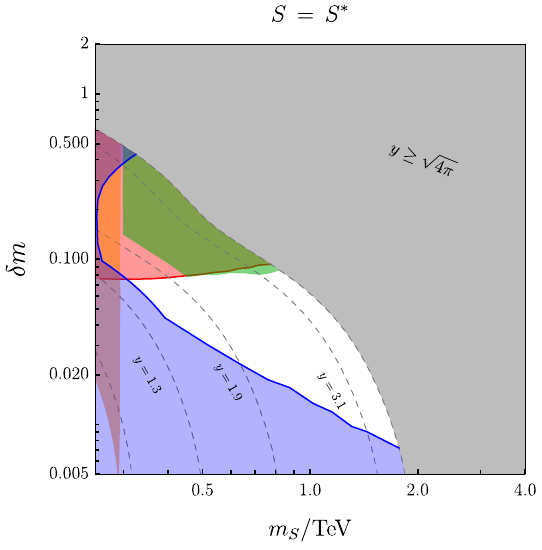}
    \caption{Relic-density-compatible parameter space in the \((m_S,\delta m)\) plane for different fixed values of the Yukawa coupling \(y\). The blue-shaded regions are excluded by direct-detection constraints, while the red-shaded regions are excluded by global SMEFT fits. The orange and green regions are excluded by Fermi-LAT and H.E.S.S. gamma-ray searches, respectively. 
    }
    \label{fig:dm_mS_SMEFT_DD_ID}
\end{figure}

We now discuss the main features of Fig.~\ref{fig:4panels_y_ms}, which will also guide our discussion of the summary plot in Fig.~\ref{fig:dm_mS_SMEFT_DD_ID}. We display the two constraints that have been the main focus of our work, direct detection and SMEFT global fits, while combining the indirect detection in the summary plots later. For moderate mass splittings, \(\delta m\gtrsim 0.5\), coannihilations involving the vector-like fermion become inefficient, and the relic abundance is determined predominantly by dark-matter pair annihilations. Since these processes are either velocity-suppressed or suppressed by powers of \(\alpha_{\rmii{QED}}\), relatively large Yukawa couplings are required to reproduce the observed relic abundance. As the mass splitting decreases, coannihilation processes become increasingly efficient, reducing the Yukawa coupling required to obtain the correct relic density. These additional channels involve the vector-like fermion and depend on the Yukawa and electroweak couplings. The black-dashed curves move consistently towards smaller $y$ values. 

The direct-detection constraints, shown by the blue-shaded regions in Fig.~\ref{fig:4panels_y_ms}, exhibit a complementary dependence on the mass splitting. In particular, the constraints become increasingly relevant as the dark-matter and mediator masses approach degeneracy. This behaviour reflects the strong dependence of the relevant matching coefficients on the mass splitting, which enhances the direct-detection rate in the near-degenerate regime. We remind the reader, however, that the direct-detection constraint carries an order-one uncertainty due to the poorly known two-proton nuclear matrix element. 

Finally, we consider the constraints from global SMEFT fits, shown by the red-shaded regions. These arise solely from the four-lepton operator, whose matching coefficient depends on the electroweak and new-physics couplings, as well as the mediator mass; see Eq.~\eqref{Cee_scalar_real}. Since the global fits constrain the corresponding SMEFT coefficient around its experimentally preferred value, the resulting bounds become more relevant as the Yukawa coupling increases. Consequently, they predominantly constrain the large-\(y\), low-mass region of the parameter space. We note that one has to regard the SMEFT constraints at the lower end of our mass range as indicative, while the EFT description becomes increasingly reliable for larger dark-sector masses.

We now turn to the summary plots in Fig.~\ref{fig:dm_mS_SMEFT_DD_ID}, where we show the parameter space in the \((m_S,\delta m)\) plane, allowing the Yukawa coupling to vary up to \(y_{\rm max}=\sqrt{4\pi}\). The dashed black contours indicate fixed values of \(y\). For \(m_S\geq 250\) GeV, these contours span the cosmologically viable parameter space for different values of the BSM coupling, which is otherwise unknown. On the lower edge of the DM mass, we find $y \simeq 0.3$.
For each fixed value of \(y\), the relic-density requirement $\Omega_{\rmii{DM}} h^2 = 0.12(12)$ determines the corresponding combinations of the dark-matter mass and mass splitting. We can then identify the regions excluded by direct detection (blue), global SMEFT fits (red), and indirect detection (green for H.E.S.S. and orange for Fermi-LAT).

As anticipated, the SMEFT global fits primarily constrain the region of large Yukawa couplings and low dark-matter masses. These constraints do not become significantly stronger as the mass splitting decreases, see Fig.~\ref{fig:Cee_real_complex}, and gradually weaken, disappearing at \(m_S \approx 850\) GeV for \(y=y_{\rm max}\). For large mass splittings, the global SMEFT fits provide stronger constraints than direct detection (blue region on top-left of the parameter space). However, direct detection also excludes a narrow region at very small mass splittings, due to the corresponding enhancement of one-loop form factors that compensate for smaller values of $y$, hence extending to dark-matter masses slightly above \(1\) TeV.

Indirect detection provides complementary constraints, particularly at the lower end of the mass range, where the Fermi-LAT limits exclude nearly all mass splittings up to \(m_S=294\) GeV. We emphasize once again that these exclusions should be interpreted as indicative, since we have not performed a dedicated recast of the Fermi-LAT analysis accounting for the distinct photon spectra of the line and continuum signals.
At higher masses, the H.E.S.S. line searches extend the indirect-detection reach to \(m_S\gtrsim 300\) GeV, yielding constraints that largely overlap with, and performs slightly better than, those from the SMEFT global fits for \(y\gtrsim 2.0\). At lower masses, the \(C_{ee}\) constraint from the SMEFT global fits is more restrictive. The SMEFT constraints also cover the narrow mass interval left unexplored between the Fermi-LAT and H.E.S.S. searches.

We now investigate how the conclusions above are modified if the
standard cosmological history is relaxed. In the right panel of
Fig.~\ref{fig:dm_mS_SMEFT_DD_ID}, we show the corresponding
relic-density-compatible parameter space for the modified cosmology
introduced in Sec.~\ref{sec:mod_cosm}, adopting
$n=2$ (kination) and $T_{\rm ref}=1~\mathrm{GeV}$. The comparison with the left
panel isolates the impact of the modified expansion rate on the
particle-physics parameters required to reproduce the observed relic
abundance.

The faster expansion rate during freeze-out reduces the efficiency of
dark-matter depletion for fixed particle-physics parameters. As a
result, reproducing the observed relic abundance requires a larger
effective cross section. In the present model, this is achieved by
moving towards larger Yukawa couplings and, for fixed $y$, towards
smaller dark-matter masses and smaller mass splittings, where the
annihilation and coannihilation processes are more efficient. This is also reflected in the minimum Yukawa coupling required at
the lower end of the mass range. For $m_S=250~\mathrm{GeV}$, we find
$y\simeq 0.75$, compared with $y\simeq 0.3$ in the standard cosmology.

The impact on the SMEFT constraints follows from the same shift in the
relic-density-compatible parameter space. Since the global SMEFT fits
preferentially constrain large Yukawa couplings and lower mediator
masses, the modified cosmology changes their overlap with the
cosmologically viable region. In particular, the larger Yukawa
couplings required by the modified expansion rate lead to a stronger SMEFT
constraint, as visible in the larger red-shaded region. A similar effect is present for indirect detection. The most relevant effects show up in direct detection, which now benefits from the interplay of larger $y$'s, smaller DM masses, and smaller mass splittings preferred by the relic density.

Overall, a substantial portion of the cosmologically viable parameter space remains unconstrained by current experimental searches, at least when one assumes a standard cosmological history, highlighting the potential for future direct- and indirect-detection experiments to probe the model. An improvement of the current SMEFT global fits can also help constrain leptophilic DM models.

\section{Conclusions}
\label{sec:conclusions}

In this work, we scrutinize leptophilic scalar dark matter in light of the latest direct-detection sensitivities and state-of-the-art global SMEFT fits, adopting the assumption of Minimal Flavour Violation (MFV) to restrict the number of independent operators.

Our contributions address several aspects of leptophilic scalar dark matter phenomenology. First, we have accounted for coannihilations with the vector-like fermion mediator in the calculation of the relic abundance. Given the electromagnetic charge of the mediator, we have included the corresponding Sommerfeld effects, which enhance and suppress the annihilation cross sections for $F\bar{F}$ and $FF$ initial states, respectively. Furthermore, we have incorporated bound-state formation effects, which, to the best of our knowledge, have not been considered in previous studies of this scenario. In the case of real scalar dark matter, these effects lead to only quite moderate corrections, inducing up to a $5\%$ shift in the dark matter mass and couplings required to reproduce the observed relic abundance. This is due to a partial cancellation between the attractive $F\bar{F}$ channel, which receives Sommerfeld enhancement and bound-state formation contributions, and the repulsive $FF$ channel, whose annihilation cross section is suppressed by the Sommerfeld factor. The near-threshold effects are more important for the complex scalar dark matter scenario, giving corrections of order $10\%$, as the repulsive channel for mediator--mediator annihilations is not present.

A further important result of our analysis is the impact of global SMEFT fits on the relic-density-compatible parameter space. We have performed the complete one-loop matching of the model onto the dimension-six Warsaw basis and evolved the resulting Wilson coefficients to the reference scale of the global fit, consistently accounting for renormalization-group effects. The resulting constraints, driven primarily by the four-lepton operator \(C_{ee}\), are complementary to those from direct and indirect detection and predominantly probe large Yukawa couplings and relatively low dark matter masses. 

As a further original contribution, we have computed the one-loop corrections to dark matter annihilation into leptons, $SS \to \ell \bar{\ell}$. These corrections retain the velocity suppression of the leading-order process. Nevertheless, they become relevant for large Yukawa couplings, where their negative contribution to the annihilation cross section leads to a non-trivial interplay with the $2 \to 3$ velocity-independent radiative process $SS \to \ell\bar{\ell}\gamma$. The combined effect of these contributions is therefore important for a consistent determination of the relic abundance in the region of parameter space with large Yukawa couplings.

We have also computed the effective operators that induce interactions between dark matter and the electromagnetic field over the full mass-ratio dependence relevant to our analysis. While effective DM-photon interactions have been addressed in previous studies, existing analytic results relevant to this model are primarily given in the limit $m_{\rmii {DM}} \ll m_{\rm med}$. Our results extend these calculations beyond the heavy-mediator limit, thereby providing a more comprehensive description of the model's electromagnetic interactions. For complex scalar dark matter, we confirm the presence of charge-radius operators, whereas in the real scalar case, the leading electromagnetic interaction is described by the Rayleigh operator. Consequently, direct detection is substantially more constraining for the complex scalar scenario, while the resulting exclusions are considerably weaker for real scalar dark matter, owing to the additional suppression of its electromagnetic interactions.  The full mass-ratio dependence is particularly relevant for small to moderate mass splittings, where the dark matter and mediator remain close enough in mass for coannihilation effects to be important, while also allowing us to extend the direct-detection analysis beyond the heavy-mediator regime. 
We further emphasize that the uncertainty in the relevant two-photon nuclear matrix element remains of order unity and therefore represents a significant theoretical uncertainty in the direct-detection limits for real scalar dark matter.

We have incorporated the latest limits from gamma-ray line searches, providing an up-to-date assessment of the experimental constraints on the model. Overall, the different probes provide complementary sensitivity across a broad range of mass splittings. For moderate mass splittings, global SMEFT fits provide competitive constraints with indirect detection, while direct detection becomes increasingly powerful as the dark matter and mediator masses approach degeneracy. In particular, for a mass splitting of around $1\%$, direct-detection bounds become the most restrictive. In the standard cosmological history, the global SMEFT fits provide competitive constraints up to dark matter masses of approximately $850~\mathrm{GeV}$, while direct detection can probe masses slightly above the TeV scale in the nearly degenerate regime.

We have also explored the impact of a modified cosmological history on the viable parameter space. A faster expansion rate during freeze-out requires larger annihilation rates to reproduce the observed relic abundance, resulting in larger Yukawa couplings for fixed dark matter and mediator masses. For the representative benchmark considered in this work, this shifts the relic-density-compatible parameter space towards larger couplings and modifies the relative coverage of the different experimental probes. 

Although our analysis assumes Minimal Flavour Violation, with three copies of the fermionic mediator, the results for the cross sections entering the relic-density calculation and for direct detection can be readily adapted to a scenario with a single mediator flavour. 

Finally, our results highlight that WIMP-like dark matter scenarios remain far from being fully explored. In particular, TeV-scale dark matter with only leptophilic interactions can still evade the most stringent constraints from collider, direct and indirect detection experiments, while global SMEFT fits provide increasingly important complementary constraints. These results illustrate that, despite the extensive experimental and theoretical efforts devoted to dark matter searches, significant regions of the cosmologically viable parameter space remain uncharted. Continued efforts to investigate these scenarios are therefore well motivated, and the next generation of experiments may offer the opportunity to probe these currently inaccessible regions.

\section*{Acknowledgement}
LT and SB are grateful to Stefano Di Noi and Riccardo Bartocci for useful discussions. LT thanks the Institute of Physics of the University of Freiburg for its hospitality.

\appendix
\renewcommand{\thesection}{\Alph{section}}
\renewcommand{\thesubsection}{\Alph{section}.\arabic{subsection}}
\renewcommand{\theequation}{\Alph{section}.\arabic{equation}}

%
\section{SMEFT matching coefficients}
\label{app:WC_scalar_DM}
In this section we list the remaining matching coefficients that are not already given in the main body of the paper in Sec.~\ref{sec:SMEFT_and_DM}. As mentioned already, we used \texttt{Matchete} \cite{Fuentes-Martin:2022jrf} for the matching of the DM model.
For later convenience we also give here the loop functions, where the mass squared ratio is $\rho= m_S^2/m_F^2$ 
\begin{align}
 \mathcal{F}_1 (\rho)   &=
\frac{
7 - 36 \rho + 45 \rho^2 - 16 \rho^3
+ 6 \rho^2 (-3 + 2 \rho)\ln(\rho)
}{
6(\rho-1)^4
} \, ,
\\
\mathcal{F}_2(\rho)
&=
\frac{
1-\rho^2+2\rho\ln(\rho)
}{
(\rho-1)^3
} \, ,
\\
\mathcal{F}_3(\rho)
&= \frac{1-6\rho + 3 \rho^2 + 2 \rho^3 - 6 \rho^2 \ln (\rho)}{6(\rho-1)^4} \, ,
\\
\mathcal{F}_4(\rho)
&= -\frac{2+3 \rho - 6 \rho^2 + \rho^3 +  6 \rho  \ln (\rho)}{6(\rho-1)^4} \, .
\end{align}
The loop functions $\mathcal{F}_i(\rho)$ have been defined as convenient combinations of the internal loop functions of \texttt{Matchete}, so to make the expressions of the matching coefficients more compact.

We introduce $\mathcal{Y}_e$ being a the lepton Yukawa matrix of the Standard Model. With the exception of $C_{ee}$, the  Wilson coefficients re the same for real or complex scalar Dark Matter. Moreover, we do not display the vanishing matching coefficients. The list of coefficients read as follows:
\begin{align}
    & C_{qq}^{(1)} \, = \,  - \dfrac{  g_{1}^4}{90 (4\pi)^2 m_{F}^2} \, , \quad C_{qu}^{(1)} \, = \,  - \dfrac{ 4 g_{1}^4}{45 (4\pi)^2 m_{F}^2} \, , \quad C_{qd}^{(1)} \, = \,   \dfrac{  2 g_{1}^4}{45 (4\pi)^2 m_{F}^2}  \, ,
    \\
    & C_{uu}^{(1)} \, =\, -   \dfrac{ 8 g_{1}^4}{45 (4\pi)^2 m_{F}^2} \, \quad C_{dd}^{(1)} \, = \, - \dfrac{2}{45(4\pi)^2 m_{F}^2} g_{1}^4  \,,
    \\
    & C_{lq}^{(1)} \, = \,   \dfrac{  g_{1}^4}{15 (4\pi)^2 m_{F}^2}    \, , \quad C_{qe}^{} \, = \, \dfrac{1}{(4\pi)^2 m_{F}^2} \left(\dfrac{2}{15} g_{1}^4 -\frac{ g_{1}^2 y^2}{36} \mathcal{F}_1(\rho)\right) \, ,
    \\
    &  C_{lu}^{} \, = \,   \dfrac{ 4 g_{1}^4}{15 (4\pi)^2 m_{F}^2}  \, , \quad   C_{ld}^{} \, = \,  - \dfrac{ 2 g_{1}^4}{15 (4\pi)^2 m_{F}^2} \, ,  \quad C_{eu}^{} \, = \, \dfrac{1}{(4\pi)^2 m_{F}^2} \left( \dfrac{8}{15} g_{1}^4 -\frac{g_{1}^2 y^2 }{9} \mathcal{F}_1(\rho)\right) \, ,
    \\
    &  C_{Hq}^{(1)} \, = \, - \dfrac{g_{1}^4}{15 (4\pi)^2 m_{F}^2} \, , \quad C_{Hu}^{} \, = \,  - \dfrac{ 4 g_{1}^4}{15 (4\pi)^2 m_{F}^2} \, , \quad  C_{Hd}^{} \, = \,   \dfrac{2 g_{1}^4}{15 (4\pi)^2 m_{F}^2} \, ,
    \\
    & C_{eB} \, = -  \dfrac{g_1 y^2 \mathcal{Y}_{e}^{}}{4 (4\pi)^2   m_{F}^2} \mathcal{F}_3(\rho) \,,  \\
    &   C_{He}^{} \, = \, \dfrac{1}{(4\pi)^2 m_{F}^2}  \left(  \frac{2}{5}g_1^4 -  \frac{g_1^2 y^2}{12} \mathcal{F}_1(\rho) \right)  \, ,  \\
    & C_{Hl}^{(1)} \, = \,  \dfrac{1}{(4\pi)^2 m_{F}^2} \left( \dfrac{1}{5} g_{1}^4  +\dfrac{y^2}{4} \mathcal{Y}_{e}^{\dagger} \mathcal{Y}_{e}  \mathcal{F}_4(\rho) \right)    \, , \quad C_{Hl}^{(3)\,} \, = \,  \dfrac{y^2 \mathcal{Y}_{e}^{\dagger} \mathcal{Y}_{e\,}}{(4\pi)^2 m_{F}^2} \mathcal{F}_4(\rho) \, ,
     \\
    & C_{H\Box} \, =  - \dfrac{g_{1}^4}{10 (4\pi)^2 m_{F}^2} \,,   \quad C_{HD} \, = \,  - \dfrac{2 g_{1}^4  }{5 (4\pi)^2 m_{F}^2}  \, .
\end{align}

%
\section{Details on direct detection}
\label{app:DD}

In this section, we provide the explicit expressions for the one-loop functions computed in this work. These functions were introduced in Sec.~\ref{sec:DD} as part of the matching coefficients of the effective DM--photon operators.

\subsection{Loop functions}
We begin with the charge-radius loop function. For the heavier lepton flavours, $\mu$ and $\tau$, the loop function $\mathcal{F}^{(a)}_\gamma(\rho,\tilde{\rho})$ entering $c_{\gamma}$ reads
\begin{align}
\mathcal{F}^{(\mu,\tau)}_\gamma(\rho,\tilde{\rho})
&=
\frac{1}
{\rho\,\lambda(1,\rho,\tilde{\rho})^{3/2}}
\Bigg\{
(1+\tilde{\rho}+2\rho)\,
\lambda(1,\rho,\tilde{\rho})^{3/2}
\ln\!\left(\tilde{\rho}^{1/2}\right)
\nonumber\\
&
+(1-\tilde{\rho})
\Bigg[
\rho(1+\tilde{\rho}-\rho)\,
\lambda(1,\rho,\tilde{\rho})^{1/2}
-
\mathcal{P}_\gamma(\rho,\tilde{\rho})
\ln\!\left(
\frac{2\tilde{\rho}^{1/2}}
{1+\tilde{\rho}-\rho+\lambda(1,\rho,\tilde{\rho})^{1/2}}
\right)
\Bigg]
\Bigg\},
\end{align}
where the Källén function is defined as
$\lambda(a,b,c)=a^2+b^2+c^2-2ab-2ac-2bc$, and the polynomial is given by
\begin{equation}
\mathcal{P}_{\gamma}(\rho,\tilde{\rho})
=
\tilde{\rho}^3
-(1+\rho)\tilde{\rho}^2
-(1+10\rho+\rho^2)\tilde{\rho}
+(1-\rho)^2(1+\rho)\,.
\end{equation}
Here, $\tilde{\rho}$ is defined in terms of the lepton flavour and its corresponding mass. The expression for the electron is
\begin{align}
\mathcal{F}^{(e)}_\gamma(\rho,\xi)
&=
\frac{\mathcal{B}(\rho,\xi)}
{16 \rho^2 (1-\rho)(1+\rho)^2}\,,
\end{align}
where the numerator is given by
\begin{align}
\mathcal{B}(\rho,\xi)
&=
12\rho(1-\rho)(1+\rho)^3\ln\xi
\nonumber\\
&
+2\rho(1+\rho)^2(5\rho^2+3)\ln\rho
-2\rho\left(5\rho^4+10\rho^3+8\rho^2-18\rho-5\right)
\ln(1-\rho)
\nonumber\\
&
+2\rho\left(5\rho^4+26\rho^3+56\rho^2+30\rho+11\right)
\ln 2
+\rho(1+\rho)
\left(8\rho^4+15\rho^3+\rho^2-23\rho-33\right).
\end{align}

Next, following the discussion of the DM interaction with a single photon, we give the expression for the Rayleigh loop function. For the heavier lepton flavours, $\mu$ and $\tau$, the loop function $\mathcal{F}^{(a)}_{\gamma\gamma}(\rho,\tilde{\rho})$ entering $c_{\gamma\gamma}$ reads
\begin{align}
\mathcal{F}_{\gamma\gamma}^{(\mu,\tau)}(\rho,\tilde{\rho})
&=
\frac{\rho^2}
{\lambda(1,\rho,\tilde{\rho})^{5/2}}
\Bigg[
\left(\rho^2+(\tilde{\rho}-1)^2-2\rho(1+\tilde{\rho})\right)
\sqrt{\lambda(1,\rho,\tilde{\rho})}
\nonumber\\
&\quad
+\mathcal{P}_{\gamma\gamma}(\rho,\tilde{\rho})
\ln\!\left(
\frac{1-\rho+\tilde{\rho}+\sqrt{\lambda(1,\rho,\tilde{\rho})}}
{2\sqrt{\tilde{\rho}}}
\right)
\Bigg],
\end{align}
where the polynomial is
\begin{equation}
\mathcal{P}_{\gamma\gamma}(\rho,\tilde{\rho})
=
\rho^3
-3\rho^2(1+\tilde{\rho})
+\rho\left(3+2\tilde{\rho}+3\tilde{\rho}^2\right)
-(1-\tilde{\rho})^2(1+\tilde{\rho})\,.
\end{equation}
The corresponding expression for the electron is
\begin{equation}
\mathcal{F}_{\gamma\gamma}^{(e)}(\rho,\xi)
=
\frac{1}{\rho}
\left[
2\ln\!\left(\frac{\xi}{(1-\rho)^2}\right)
-\operatorname{Li}_2\!\left(
1-\frac{\xi}{(\rho-1)^2}
\right)
\right],
\end{equation}
where $\operatorname{Li}_2(x)$ denotes the dilogarithm function.

Finally, we note that a more refined direct-detection analysis should also account for renormalization-group (RG) effects. At dimension six, we find that $c_{\gamma\gamma}$ does not mix with other operators. For complex DM, a non-vanishing $c_\gamma$ can instead mix into the operator
$(S^* \overset{\leftrightarrow}{\partial_\mu} S)(\bar{\ell}\gamma^\mu \ell)$, as reported in Ref.~\cite{Kavanagh:2018xeh}.

\subsection{DM--nucleon cross section for the charge-radius operator}

We derive the recoil cross section induced by the single-photon operator and mostly follow the strategy outlined in Ref.~\cite{Ibarra:2015fqa}. We consider scattering off a nucleus rather than an individual nucleon and then finally express the result in terms of the experimentally quoted DM--nucleon cross section.
The nuclear mass is denoted by $m_A$, and the corresponding reduced mass is
\begin{equation}
    \mu_{\rm red}^{(A)}
    =
    \frac{m_S m_A}{m_S+m_A}\,.
\end{equation}

We write the operator in terms of a generic mass scale $\Lambda$, which will later be set to $m_S$:
\begin{equation}
    \mathcal{O}_\gamma
    =
    i \frac{c_\gamma}{m_S^2}
    \partial^\mu S \partial^\nu S^* F_{\mu \nu} \, .
\end{equation}

The amplitude for $S A \to S A$, obtained from this vertex and the effective nucleus--photon coupling, is
\begin{eqnarray}
    i \mathcal{M}^{\gamma}_{S A \to S A}
    =
    \left(
    -i \frac{c_\gamma}{2 m_S^2}
    q^2 (p+p')^\mu
    \right)
    \frac{-i g_{\mu \nu}}
    {q^2+i \epsilon}
    \langle i e J_A^\nu \rangle \, ,
\end{eqnarray}
where the matrix element of the nuclear electromagnetic current is
\begin{eqnarray}
    \langle A(k') | J_A^\nu | A(k) \rangle
    =
    Z (k+k')^\nu
    F_{\rm SI}(q^2)\, .
\end{eqnarray}
Here, $F_{\rm SI}(q^2)$ is the spin-independent nuclear form factor describing the coherent coupling of the photon to the protons, and $Z$ is the atomic number. The factors of the momentum transfer from the effective vertex cancel against the photon propagator, yielding
\begin{equation}
    i \mathcal{M}^{\gamma}_{S A \to S A}
    =
    - i \frac{c_\gamma}{2 m_S^2}
    Z e
    F_{\rm SI}(q^2)
    (p+p') \cdot (k+k') \, .
\end{equation}

For the spin-independent form factor, we use the Helm parametrisation,
\begin{equation}
F_{\mathrm{SI}}^2(q)
=
\left[
\frac{3\,j_1(qR_1)}{qR_1}
\right]^2
e^{-q^2 s^2},
\end{equation}
with
\begin{equation}
R_1 = \sqrt{R^2 - 5s^2},
\qquad
R = 1.2\,A^{1/3}\,\mathrm{fm},
\qquad
s \simeq 0.9\,\mathrm{fm},
\end{equation}
where $A$ is the atomic mass number and $j_1$ is the spherical Bessel function,
\begin{equation}
j_1(x)
=
\frac{\sin x}{x^2}
-
\frac{\cos x}{x}.
\end{equation}

We now simplify the amplitude using non-relativistic kinematics. The four-momenta are
\begin{align}
   & (p+p')^{\mu}
   =
   (E_{p} + E_{p'},\bm{p}+\bm{p}')
   \simeq
   \left(
   2 m_S + m_S v^2 - \bm{v} \cdot \bm{q}
   + \frac{\bm{q}^2}{2 m_S},
   2 m_S \bm{v}-\bm{q}
   \right) ,
   \\
   & (k+k')^{\mu}
   =
   (E_k + E_{k'},\bm{k}+\bm{k}')
   \simeq
   \left(
   2 m_A + \frac{\bm{q}^2}{2 m_A},
   \bm{q}
   \right) .
\end{align}
Retaining terms up to $\mathcal{O}(v^2)$ and $\mathcal{O}(q^2)$, we obtain
\begin{eqnarray}
    (p+p') \cdot (k+k')
    =
    4 m_S m_A
    +
    2 m_S m_A v^2
    -
    2 (m_S + m_A)\,\bm{v}\cdot\bm{q}
    +
    \frac{m_S}{m_A}\bm{q}^2
    +
    \bm{q}^2
    \left(
    1+\frac{m_A}{m_S}
    \right).
    \nonumber \\
    \phantom{s}
\end{eqnarray}
Using
\begin{eqnarray}
    E_R
    =
    \frac{\bm{q}^2}{2 m_A},
    \qquad
    \bm{v}\cdot\bm{q}
    =
    \frac{\bm{q}^2}
    {2 \mu_{\rm red}^{(A)}},
\end{eqnarray}
we can express the result in terms of $v$ and $E_R$:
\begin{eqnarray}
     (p+p') \cdot (k+k')
     =
     4 m_S m_A
     +
     2 m_S m_A v^2
     -
     2 m_A E_R \, .
\end{eqnarray}
The momentum dependence of the Helm form factor can likewise be expressed in terms of the recoil energy,
\begin{equation}
    |\bm q|
    =
    \sqrt{2 m_A E_R}\, .
\end{equation}

The squared matrix element is therefore
\begin{eqnarray}
|\mathcal{M}^{\gamma}_{S A \to S A}|^2
&=&
\left(
\frac{eZ c_\gamma}
{2 m_S^2}
\right)^2
F_{\rm SI}^2(E_R)
\left(
4 m_S m_A
+
2 m_S m_A v^2
-
2 m_A E_R
\right)^2
\\
&=&
\left(
\frac{eZ c_\gamma}
{m_S^2}
\right)^2
F_{\rm SI}^2(E_R)
\,4 m_S^2 m_A^2
\left(
1+\frac{v^2}{2}
-\frac{E_R}{2m_S}
\right)^2 .
\end{eqnarray}

For typical dark matter velocities, $v^2 \sim 10^{-6}$, and recoil energies, $E_R \sim {\rm few}\ {\rm keV}-50\,{\rm keV}$, we have $E_R/m_S \ll 1$. Both corrections are therefore negligible, and we retain only the leading term:
\begin{eqnarray}
|\mathcal{M}^{\gamma}_{S A \to S A}|^2
\simeq
\left(
\frac{eZ c_\gamma}
{m_S^2}
\right)^2
F_{\rm SI}^2(E_R)
\,4 m_S^2 m_A^2 \, .
\end{eqnarray}

The differential cross section is then
\begin{eqnarray}
\frac{d \sigma}{d E_R}
=
\frac{e^2 Z^2 c_\gamma^2}{8 \pi v^2}
\frac{F_{\rm SI}^2(E_R) m_A}{m_S^4} \, .
\end{eqnarray}
In the zero-momentum-transfer limit, $F_{\rm SI}^2(E_R)\to 1$, and integrating up to the maximum recoil energy,
\begin{eqnarray}
E_R^{\rm max}
=
2 \left(\mu_{\rm red}^{(A)}\right)^2
\frac{v^2}{m_A},
\end{eqnarray}
gives the total nuclear cross section
\begin{eqnarray}
\sigma_{\rm SI}^{(A)}
=
\frac{e^2 Z^2 c_\gamma^2}{4 \pi m_S^4}
\left(\mu_{\rm red}^{(A)}\right)^2 \, .
\label{our_SI_single_photon}
\end{eqnarray}
This approximation is sufficient for comparing the predicted cross section with the exclusion limits reported by direct-detection experiments. A calculation of the full recoil spectrum and event rate would instead require retaining the finite-momentum-transfer dependence encoded in the Helm form factor.

To compare with experimental limits, we express the result in terms of the effective proton coupling $f_p$, defined through the standard spin-independent nuclear cross section,
\begin{eqnarray}
\sigma^{(A)}_{\rm SI}
=
\frac{(\mu_{\rm red}^{(A)})^2}{\pi}
\left[
Zf_p+(A-Z)f_n
\right]^2 .
\end{eqnarray}
Matching this expression to Eq.~\eqref{our_SI_single_photon}, and retaining the leading spin-independent contribution from the electromagnetic charge density, which is proportional to the proton number $Z$, we identify
\begin{eqnarray}
f_n=0,
\qquad
f_p=\frac{e c_\gamma}{2m_S^2}.
\end{eqnarray}
The corresponding proton cross section is
\begin{eqnarray}
\sigma^{p}_{\rm SI}
=
\frac{(\mu_{\rm red}^{(N)})^2}{\pi}f_p^2
=
\frac{\alpha_{\rmii{QED}} c_\gamma^2}{m_S^4}
(\mu_{\rm red}^{(N)})^2.
\end{eqnarray}

In our numerical analysis, we compare this cross section with the mass-dependent upper limits reported by the LZ collaboration~\cite{LZ:2024zvo} (Fig.~5). These limits assume equal couplings to protons and neutrons, $f_p=f_n$. Since the interaction considered here is proton-only at the one-body level, with $f_n=0$, the comparison should be understood as approximate. A dedicated recast accounting for the isospin dependence of the xenon response would be required to obtain precise exclusion limits for this interaction.

%

\section{Cross sections for the relic density}
\label{app:cross_sections_relic}
Here we collect the   cross sections of the $2 \to 2$ processes as needed to estimate the DM energy density. We refrain from listing the expressions with finite mass ratios, $m_{\rmii{SM}}^2/m_S^2$ and $m_{\rmii{SM}}^2/m_F^2$ as these are typically small in the DM mass range explored in our work. We provide the leading velocity contribution in the following, hence making the list of cross section compact.  In our numerical extraction of the cross section we keep both the mass ratios and velocity suppressed contribution up to order $\vrel^2$, and consider the Sommerfeld factors accordingly for the velocity independent and velocity dependent terms of the cross sections. 

The coannihilation processes with one fermion mediator and one scalar read 
\begin{align}
      \sigma(S F_a \to e_a \gamma ) \vrel &=  \frac{\alpha_{\rmii{QED}} y^2  Y_F Y_e \,       m_S}{8 \pi \,m_F^2(m_F+m_S)}  \label{g}\, ,
      \\[3pt]
      \sigma(S F_a \to e_a Z) \vrel &=  \frac{ \alpha_{\rmii{QED}} \tan\theta_{\scriptscriptstyle W}^2 y^2  Y_F Y_e \,       m_S }{8\pi \, m_F^2 (m_F+m_S)} \, ,
\end{align}
whereas the mediator-antimediator  annihilations are 
\begin{align}
      \sigma (F_a \bar{F}_a \to \gamma \gamma) \vrel &=  \frac{ \pi \alpha_{\rmii{QED}}^2 Y_F^4 }{m_F^2}  \, , 
       \\[3pt]
      \sigma (F_a \bar{F}_a \to ZZ) \vrel  &= \frac{ \pi \alpha_{\rmii{QED}}^2 \tan^4\theta_{\scriptscriptstyle W} Y_F^4 }{m_F^2}  \, ,
      \\[3pt]
      \sigma (F_a \bar{F}_a \to \gamma Z) \vrel &=  \frac{\, \pi \alpha_{\rmii{QED}}^2 Y_F^4 \tan^2\theta_{\scriptscriptstyle W} }{ \,m_F^2} \, ,
        \\[3pt]
         \sigma (F_a \bar{F}_a \to W^+ W^-) \vrel &= \frac{\pi \alpha_{\rmii{QED}}^2  Y_F^4}{16  c_{\scriptscriptstyle W}^4 m_F^2}\,,
          \\[3pt]
         \sigma (F_a \bar{F}_a \to  Z \, h) \vrel &=0\,,
          \\[3pt]
           \sigma (F_a \bar{F}_a \to e_a \bar{e}_a) \vrel &= \dfrac{1}{32 \pi} \left[ \dfrac{y^4 \, m_F^2}{(m_F^2 +m_S^2)^2}+ \dfrac{16\pi^2 \alpha_{\rmii{QED}}^2}{m_F^2 c_{\scriptscriptstyle W}^4}- \dfrac{4 \pi \alpha_{\rmii{QED}}\, y^2}{c_{\scriptscriptstyle W}^2 (m_F^2 +m_S^2)}\right]\,,
\\[3pt]
           \sigma  (F_a \bar{F}_a \to e_b \bar{e}_b) \vrel &= \dfrac{\pi \alpha_{\rmii{QED}}^2}{2\,c_W^4 m_F^2}\,, \quad a \neq b
\\[3pt]
           \sigma  (F_a \bar{F}_b \to e_a \bar{e}_b) \vrel &= \dfrac{ y^4\,m_F^2}{32 \pi (m_F^2 +m_S^2)^2}\,, \quad a \neq b
           \\[3pt]
           \sigma  (F_a \bar{F}_a \to \ell_b \bar{\ell}_b) \vrel &= \dfrac{ \pi \alpha_{\rmii{QED}}^2}{8\,c_{\scriptscriptstyle W}^4 m_F^2}\,,
           \label{FFbar_to_LHlep}
           \\[3pt]
           \sigma  (F_a \bar{F}_a \to \nu_b \bar{\nu}_b) \vrel &= \dfrac{\pi \alpha_{\rmii{QED}}^2}{8\,c_{\scriptscriptstyle W}^4 m_F^2}\,,
           \\[3pt]
           \sigma  (F_a \bar{F}_a \to u_b \bar{u}_b) \vrel &= \dfrac{17 \pi N_{\rmi{c}} \,  \alpha_{\rmii{QED}}^2 }{72 c_{\scriptscriptstyle W}^4 m_F^2}\,,
           \\[3pt]
           \sigma  (F_a \bar{F}_a \to d_b \bar{d}_b) \vrel &= \dfrac{5 \pi N_{\rmi{c}} \, \alpha_{\rmii{QED}}^2 }{72 c_{\scriptscriptstyle W}^4 m_F^2}\,,
           \label{FFbar_to_RHquarks}
\end{align}
We recall the notation for the SM fermions: $e$ and $\ell$ denote right-handed and left-handed charged leptons, respectively, while $\nu$ denotes left-handed neutrinos. The fields $u$ and $d$ represent up-type and down-type quarks, respectively. Each process listed above is given for a single flavour of the vector-like fermion $F$ and the corresponding SM fermions, with cases involving distinct flavours ($a\neq b$) indicated explicitly; $N_{\rmi{c}}=3$ is the number of colours for SM strong interactions.  The processes in Eqs.~\eqref{FFbar_to_LHlep}--\eqref{FFbar_to_RHquarks} are mediated by $s$-channel photon or $Z$-boson exchange, in analogy with the Drell--Yan process.

Next, we write the fermion-fermion annihilation cross sections, which are relevant for the real scalar DM, and are found to be 
\begin{align}
\sigma  (F_a F_a \to e_a e_a)  \vrel  &=  \dfrac{y^4 \,m_F^2}{32 \pi (m_F^2 + m_S^2)^2}\, ,
\\[3pt]
\sigma (F_a F_b \to e_a e_b) \vrel  &=  \dfrac{y^4 \,m_F^2}{32 \pi (m_F^2 + m_S^2)^2}\, , \quad a \neq b \, .
\end{align}

The resulting combinations of the cross section for the mediator-antimediator and mediator-mediator (and its complex conjugate process) read 
\begin{align}
        \sigma_{F\bar{F}} \vrel &= \sigma   (F_ a \bar{F}_a \to e_a \bar{e}_a)  \vrel+ 2 \sigma (F_a \bar{F}_a \to e_b \bar{e}_b) \vrel  +\sigma   (F_a \bar{F}_b \to e_a \bar{e}_b) \vrel
        \nonumber \\
       & + \sigma   (F_ a \bar{F}_a \to VV)  \vrel + \sigma   (F_ a \bar{F}_a \to Z h) \vrel 
         \nonumber \\
       & + \sigma  (F_a \bar{F}_a \to \ell_b \bar{\ell}_b) \vrel  +  \sigma  (F_a \bar{F}_a \to \nu_b \bar{\nu}_b) \vrel 
        \nonumber \\
       & + \sigma  (F_a \bar{F}_a \to u_b \bar{u}_b) \vrel +\sigma  (F_a \bar{F}_a \to d_b \bar{d}_b) \vrel
       \, , 
       \\
        \sigma_{FF} \vrel &= \sigma    (F_a F_a \to e_a e_a) \vrel + 2 \sigma  (F_a F_b \to e_a e_b)\vrel \, ,
    \end{align}
    where flavor indices are not summed and $a \neq b$, as for they are taken into account by the factor $n_f$ in Eq.~\eqref{eff_Xsection_coann_real_S}, and the combinations of vector bosons in the final state are ${VV}=(\gamma \gamma, ZZ, \gamma Z, W^+ W^-)$.

As a last ingredient for the extraction of the relic density in the complex DM case, let us write down the corresponding equation for the effective cross section, which reads 
\begin{align}
     \sigma_{\textrm{eff}} \vrel &= \left[ 2 \sigma_{S S^*} \vrel +  \frac{4 g_F}{g_S} e^{- \frac{\Delta m}{T}}  \left( 1+\delta m \right)^{\frac{3}{2}} \sigma_{S \bar{F}} \vrel  \right.
    \nonumber
    \\
    &\left. +  \frac{ 2 g_F^2 }{g_S^2} e^{-2  \frac{\Delta m}{T}} \left( 1+\delta m \right)^{3}  \sigma_{F \bar{F}} \vrel \right]
    \frac{n_f}{\left[ 2+ \frac{2 n_f  g_F}{g_S} e^{- \frac{\Delta m }{T}} \left( 1+\delta m \right)^{\frac{3}{2}} \right]^2} \, .
    \nonumber
    \\
    \phantom{s}
\label{eff_Xsection_coann_complex_S}
\end{align}

\bibliographystyle{hieeetr}
\bibliography{paper.bib}

@article{Arina:2025zpi,
    author = "Arina, Chiara and others",
    title = "{t-channel dark matter models {\textendash} a whitepaper}",
    eprint = "2504.10597",
    archivePrefix = "arXiv",
    primaryClass = "hep-ph",
    reportNumber = "CERN-LPCC-2025-001, IRMP-CP3-25-07, TTK-25-07",
    doi = "10.1140/epjc/s10052-025-14635-7",
    journal = "Eur. Phys. J. C",
    volume = "85",
    number = "9",
    pages = "975",
    year = "2025"
}

@article{HESS:2026ila,
    author = "Aharonian, F. and others",
    collaboration = "H.E.S.S.",
    title = "{Search for Gamma-Ray Spectral Lines from Dark Matter Annihilation with the H.E.S.S. Inner Galaxy Survey}",
    eprint = "2608.07234",
    archivePrefix = "arXiv",
    primaryClass = "astro-ph.HE",
    doi = "10.1103/d8tj-55kc",
    journal = "Phys. Rev. Lett.",
    volume = "137",
    number = "9",
    pages = "091002",
    year = "2026"
}

@article{Hisano:2004ds,
    author = "Hisano, Junji and Matsumoto, Shigeki. and Nojiri, Mihoko M. and Saito, Osamu",
    title = "{Non-perturbative effect on dark matter annihilation and gamma ray signature from galactic center}",
    eprint = "hep-ph/0412403",
    archivePrefix = "arXiv",
    reportNumber = "ICRR-REPORT-513-2004-11, YITP-04-73",
    doi = "10.1103/PhysRevD.71.063528",
    journal = "Phys. Rev. D",
    volume = "71",
    pages = "063528",
    year = "2005"
}

@article{Petraki:2015hla,
    author = "Petraki, Kalliopi and Postma, Marieke and Wiechers, Michael",
    title = "{Dark-matter bound states from Feynman diagrams}",
    eprint = "1505.00109",
    archivePrefix = "arXiv",
    primaryClass = "hep-ph",
    reportNumber = "NIKHEF-2015-013",
    doi = "10.1007/JHEP06(2015)128",
    journal = "JHEP",
    volume = "06",
    pages = "128",
    year = "2015"
}

@article{Bergstrom:1988fp,
    author = "Bergstrom, L. and Snellman, H.",
    title = "{Observable Monochromatic Photons From Cosmic Photino Annihilation}",
    doi = "10.1103/PhysRevD.37.3737",
    journal = "Phys. Rev. D",
    volume = "37",
    pages = "3737--3741",
    year = "1988"
}

@article{Rudaz:1986db,
    author = "Rudaz, Serge",
    title = "{Cosmic Production of Quarkonium?}",
    reportNumber = "UMN-TH-549/86",
    doi = "10.1103/PhysRevLett.56.2128",
    journal = "Phys. Rev. Lett.",
    volume = "56",
    pages = "2128",
    year = "1986"
}

@article{Srednicki:1985sf,
    author = "Srednicki, Mark and Theisen, Stefan and Silk, Joseph",
    title = "{Cosmic Quarkonium: A Probe of Dark Matter}",
    reportNumber = "PRINT-85-0916 (UC,SANTA-BARBARA)",
    doi = "10.1103/PhysRevLett.56.263",
    journal = "Phys. Rev. Lett.",
    volume = "56",
    pages = "263",
    year = "1986",
    note = "[Erratum: Phys.Rev.Lett. 56, 1883 (1986)]"
}

@article{Petraki:2016cnz,
    author = "Petraki, Kalliopi and Postma, Marieke and de Vries, Jordy",
    title = "{Radiative bound-state-formation cross-sections for dark matter interacting via a Yukawa potential}",
    eprint = "1611.01394",
    archivePrefix = "arXiv",
    primaryClass = "hep-ph",
    doi = "10.1007/JHEP04(2017)077",
    journal = "JHEP",
    volume = "04",
    pages = "077",
    year = "2017"
}

@article{HESS:2018cbt,
    author = "Abdallah, H. and others",
    collaboration = "HESS",
    title = "{Search for $\gamma$-Ray Line Signals from Dark Matter Annihilations in the Inner Galactic Halo from 10 Years of Observations with H.E.S.S.}",
    eprint = "1805.05741",
    archivePrefix = "arXiv",
    primaryClass = "astro-ph.HE",
    doi = "10.1103/PhysRevLett.120.201101",
    journal = "Phys. Rev. Lett.",
    volume = "120",
    number = "20",
    pages = "201101",
    year = "2018"
}

@article{Fermi-LAT:2015kyq,
    author = "Ackermann, M. and others",
    collaboration = "Fermi-LAT",
    title = "{Updated search for spectral lines from Galactic dark matter interactions with pass 8 data from the Fermi Large Area Telescope}",
    eprint = "1506.00013",
    archivePrefix = "arXiv",
    primaryClass = "astro-ph.HE",
    reportNumber = "FERMILAB-PUB-15-673-AE",
    doi = "10.1103/PhysRevD.91.122002",
    journal = "Phys. Rev. D",
    volume = "91",
    number = "12",
    pages = "122002",
    year = "2015"
}

@article{Biondini:2023ksj,
    author = "Biondini, Simone and Bollig, Julian and Vogl, Stefan",
    title = "{Indirect detection of dark matter with (pseudo)-scalar interactions}",
    eprint = "2308.14594",
    archivePrefix = "arXiv",
    primaryClass = "hep-ph",
    doi = "10.1007/JHEP04(2024)050",
    journal = "JHEP",
    volume = "04",
    pages = "050",
    year = "2024"
}

@article{Garny:2021qsr,
    author = "Garny, Mathias and Heisig, Jan",
    title = "{Bound-state effects on dark matter coannihilation: Pushing the boundaries of conversion-driven freeze-out}",
    eprint = "2112.01499",
    archivePrefix = "arXiv",
    primaryClass = "hep-ph",
    reportNumber = "TUM-HEP 1379/21, TTK-21-52",
    doi = "10.1103/PhysRevD.105.055004",
    journal = "Phys. Rev. D",
    volume = "105",
    number = "5",
    pages = "055004",
    year = "2022"
}

@article{Detmold:2014qqa,
    author = "Detmold, William and McCullough, Matthew and Pochinsky, Andrew",
    title = "{Dark Nuclei I: Cosmology and Indirect Detection}",
    eprint = "1406.2276",
    archivePrefix = "arXiv",
    primaryClass = "hep-ph",
    reportNumber = "MIT-CTP-4554",
    doi = "10.1103/PhysRevD.90.115013",
    journal = "Phys. Rev. D",
    volume = "90",
    number = "11",
    pages = "115013",
    year = "2014"
}

@article{Feng:2010zp,
    author = "Feng, Jonathan L. and Kaplinghat, Manoj and Yu, Hai-Bo",
    title = "{Sommerfeld Enhancements for Thermal Relic Dark Matter}",
    eprint = "1005.4678",
    archivePrefix = "arXiv",
    primaryClass = "hep-ph",
    reportNumber = "UCI-TR-2010-06",
    doi = "10.1103/PhysRevD.82.083525",
    journal = "Phys. Rev. D",
    volume = "82",
    pages = "083525",
    year = "2010"
}

@article{Cahill-Rowley:2015aea,
    author = "Cahill-Rowley, Matthew and El Hedri, Sonia and Shepherd, William and Walker, Devin G. E.",
    title = "{Perturbative Unitarity Constraints on Charged/Colored Portals}",
    eprint = "1501.03153",
    archivePrefix = "arXiv",
    primaryClass = "hep-ph",
    reportNumber = "SLAC-PUB-16190, MITP-14-104, SCIPP-15-01",
    doi = "10.1016/j.dark.2018.04.003",
    journal = "Phys. Dark Univ.",
    volume = "22",
    pages = "48--59",
    year = "2018"
}

@article{Arina:2020udz,
    author = "Arina, Chiara and Fuks, Benjamin and Mantani, Luca",
    title = "{A universal framework for t-channel dark matter models}",
    eprint = "2001.05024",
    archivePrefix = "arXiv",
    primaryClass = "hep-ph",
    reportNumber = "CP3-20-01, MCNET-20-01",
    doi = "10.1140/epjc/s10052-020-7933-7",
    journal = "Eur. Phys. J. C",
    volume = "80",
    number = "5",
    pages = "409",
    year = "2020"
}

@article{Buchmuller:1985jz,
    author = "Buchmuller, W. and Wyler, D.",
    title = "{Effective Lagrangian Analysis of New Interactions and Flavor Conservation}",
    reportNumber = "CERN-TH-4254/85",
    doi = "10.1016/0550-3213(86)90262-2",
    journal = "Nucl. Phys. B",
    volume = "268",
    pages = "621--653",
    year = "1986"
}

@article{Ibarra:2015fqa,
    author = "Ibarra, Alejandro and Wild, Sebastian",
    title = "{Dirac dark matter with a charged mediator: a comprehensive one-loop analysis of the direct detection phenomenology}",
    eprint = "1503.03382",
    archivePrefix = "arXiv",
    primaryClass = "hep-ph",
    doi = "10.1088/1475-7516/2015/05/047",
    journal = "JCAP",
    volume = "05",
    pages = "047",
    year = "2015"
}

@article{Bartocci:2023nvp,
    author = {Bartocci, Riccardo and Biek\"otter, Anke and Hurth, Tobias},
    title = "{A global analysis of the SMEFT under the minimal MFV assumption}",
    eprint = "2311.04963",
    archivePrefix = "arXiv",
    primaryClass = "hep-ph",
    reportNumber = "MITP/23-063",
    doi = "10.1007/JHEP05(2024)074",
    journal = "JHEP",
    volume = "05",
    pages = "074",
    year = "2024"
}

@article{Bartocci:2024fmm,
    author = {Bartocci, Riccardo and Biek\"otter, Anke and Hurth, Tobias},
    title = "{Renormalisation group evolution effects on global SMEFT analyses}",
    eprint = "2412.09674",
    archivePrefix = "arXiv",
    primaryClass = "hep-ph",
    reportNumber = "MITP-24-087",
    doi = "10.1007/JHEP05(2025)203",
    journal = "JHEP",
    volume = "05",
    pages = "203",
    year = "2025"
}

@article{Giani:2023gfq,
    author = "Giani, Tommaso and Magni, Giacomo and Rojo, Juan",
    title = "{SMEFiT: a flexible toolbox for global interpretations of particle physics data with effective field theories}",
    eprint = "2302.06660",
    archivePrefix = "arXiv",
    primaryClass = "hep-ph",
    reportNumber = "Nikhef-2022-023",
    doi = "10.1140/epjc/s10052-023-11534-7",
    journal = "Eur. Phys. J. C",
    volume = "83",
    number = "5",
    pages = "393",
    year = "2023"
}

@article{deBlas:2025xhe,
    author = "de Blas, Jorge and Goncalves, Angelica and Miralles, V{\'\i}ctor and Reina, Laura and Silvestrini, Luca and Valli, Mauro",
    title = "{Constraining new physics effective interactions via a global fit of electroweak, Drell-Yan, Higgs, top, and flavour observables}",
    eprint = "2507.06191",
    archivePrefix = "arXiv",
    primaryClass = "hep-ph",
    doi = "10.1007/JHEP03(2026)013",
    journal = "JHEP",
    volume = "03",
    pages = "013",
    year = "2026"
}

@article{Bertone:2016nfn,
    author = "Bertone, Gianfranco and Hooper, Dan",
    title = "{History of dark matter}",
    eprint = "1605.04909",
    archivePrefix = "arXiv",
    primaryClass = "astro-ph.CO",
    reportNumber = "FERMILAB-PUB-16-157-A",
    doi = "10.1103/RevModPhys.90.045002",
    journal = "Rev. Mod. Phys.",
    volume = "90",
    number = "4",
    pages = "045002",
    year = "2018"
}

@article{Biondini:2025gpg,
    author = "Biondini, Simone and Tiberi, Lorenzo and Panella, Orlando",
    title = "{Connecting t-channel dark matter models to the Standard Model Effective Field Theory}",
    eprint = "2507.00925",
    archivePrefix = "arXiv",
    primaryClass = "hep-ph",
    doi = "10.1007/JHEP10(2025)060",
    journal = "JHEP",
    volume = "10",
    pages = "060",
    year = "2025"
}

@article{Olgoso:2025jot,
    author = "Olgoso, Pablo and Paradisi, Paride and Selimovic, Nudzeim",
    title = "{The dark side of a Tera-Z factory}",
    eprint = "2507.17803",
    archivePrefix = "arXiv",
    primaryClass = "hep-ph",
    doi = "10.1007/JHEP04(2026)106",
    journal = "JHEP",
    volume = "04",
    pages = "106",
    year = "2026"
}

@article{Alloul:2013bka,
    author = "Alloul, Adam and Christensen, Neil D. and Degrande, C{\'e}line and Duhr, Claude and Fuks, Benjamin",
    title = "{FeynRules  2.0 - A complete toolbox for tree-level phenomenology}",
    eprint = "1310.1921",
    archivePrefix = "arXiv",
    primaryClass = "hep-ph",
    reportNumber = "CERN-PH-TH-2013-239, MCNET-13-14, IPPP-13-71, DCPT-13-142, PITT-PACC-1308",
    doi = "10.1016/j.cpc.2014.04.012",
    journal = "Comput. Phys. Commun.",
    volume = "185",
    pages = "2250--2300",
    year = "2014"
}

@article{Fuentes-Martin:2022jrf,
    author = {Fuentes-Mart{\'\i}n, Javier and K{\"o}nig, Matthias and Pag{\`e}s, Julie and Thomsen, Anders Eller and Wilsch, Felix},
    title = "{A proof of concept for matchete: an automated tool for matching effective theories}",
    eprint = "2212.04510",
    archivePrefix = "arXiv",
    primaryClass = "hep-ph",
    reportNumber = "MITP-22-105, TUM-HEP-1443/22, ZU-TH-58/22",
    doi = "10.1140/epjc/s10052-023-11726-1",
    journal = "Eur. Phys. J. C",
    volume = "83",
    number = "7",
    pages = "662",
    year = "2023"
}

@article{Biondini:2023zcz,
    author = "Biondini, Simone and Brambilla, Nora and Qerimi, Gramos and Vairo, Antonio",
    title = "{Effective field theories for dark matter pairs in the early universe: cross sections and widths}",
    eprint = "2304.00113",
    archivePrefix = "arXiv",
    primaryClass = "hep-ph",
    doi = "10.1007/JHEP07(2023)006",
    journal = "JHEP",
    volume = "07",
    pages = "006",
    year = "2023"
}

@article{Ovanesyan:2014fha,
    author = "Ovanesyan, Grigory and Vecchi, Luca",
    title = "{Direct detection of dark matter polarizability}",
    eprint = "1410.0601",
    archivePrefix = "arXiv",
    primaryClass = "hep-ph",
    reportNumber = "ACFI-T14-19",
    doi = "10.1007/JHEP07(2015)128",
    journal = "JHEP",
    volume = "07",
    pages = "128",
    year = "2015"
}

@article{Frandsen:2012db,
    author = "Frandsen, Mads T. and Haisch, Ulrich and Kahlhoefer, Felix and Mertsch, Philipp and Schmidt-Hoberg, Kai",
    title = "{Loop-induced dark matter direct detection signals from gamma-ray lines}",
    eprint = "1207.3971",
    archivePrefix = "arXiv",
    primaryClass = "hep-ph",
    reportNumber = "OUTP-12-12-P",
    doi = "10.1088/1475-7516/2012/10/033",
    journal = "JCAP",
    volume = "10",
    pages = "033",
    year = "2012"
}

@article{Sahni:1999gb,
    author = "Sahni, Varun and Starobinsky, Alexei A.",
    title = "{The Case for a positive cosmological Lambda term}",
    eprint = "astro-ph/9904398",
    archivePrefix = "arXiv",
    reportNumber = "IUCAA-25-2000",
    doi = "10.1142/S0218271800000542",
    journal = "Int. J. Mod. Phys. D",
    volume = "9",
    pages = "373--444",
    year = "2000"
}

@article{Caldwell:1997ii,
    author = "Caldwell, R. R. and Dave, Rahul and Steinhardt, Paul J.",
    title = "{Cosmological imprint of an energy component with general equation of state}",
    eprint = "astro-ph/9708069",
    archivePrefix = "arXiv",
    doi = "10.1103/PhysRevLett.80.1582",
    journal = "Phys. Rev. Lett.",
    volume = "80",
    pages = "1582--1585",
    year = "1998"
}

@article{Biondini:2023hek,
    author = "Biondini, Simone",
    title = "{Interplay between improved interaction rates and modified cosmological histories for dark matter}",
    eprint = "2309.00323",
    archivePrefix = "arXiv",
    primaryClass = "hep-ph",
    doi = "10.3389/fphy.2023.1285986",
    journal = "Front. in Phys.",
    volume = "11",
    pages = "1285986",
    year = "2023"
}

@article{DEramo:2017gpl,
    author = "D'Eramo, Francesco and Fernandez, Nicolas and Profumo, Stefano",
    title = "{When the Universe Expands Too Fast: Relentless Dark Matter}",
    eprint = "1703.04793",
    archivePrefix = "arXiv",
    primaryClass = "hep-ph",
    reportNumber = "SCIPP-17-02",
    doi = "10.1088/1475-7516/2017/05/012",
    journal = "JCAP",
    volume = "05",
    pages = "012",
    year = "2017"
}

@article{Shtabovenko:2020gxv,
  author = {Shtabovenko, Vladyslav and Mertig, Rolf and Orellana, Frederik},
  title = {FeynCalc 9.3: New features and improvements},
  journal = {Comput. Phys. Commun.},
  volume = {256},
  pages = {107478},
  year = {2020},
  eprint = {2001.04407},
  archivePrefix = {arXiv},
  primaryClass = {hep-ph},
  doi = {10.1016/j.cpc.2020.107478}
}

@article{LZ:2024zvo,
    author = "Aalbers, J. and others",
    collaboration = "LZ",
    title = "{Dark Matter Search Results from 4.2{\,}{\,}Tonne-Years of Exposure of the LUX-ZEPLIN (LZ) Experiment}",
    eprint = "2410.17036",
    archivePrefix = "arXiv",
    primaryClass = "hep-ex",
    reportNumber = "FERMILAB-PUB-24-0796-V",
    doi = "10.1103/4dyc-z8zf",
    journal = "Phys. Rev. Lett.",
    volume = "135",
    number = "1",
    pages = "011802",
    year = "2025"
}

@article{Iengo:2009ni,
    author = "Iengo, Roberto",
    title = "{Sommerfeld enhancement: General results from field theory diagrams}",
    eprint = "0902.0688",
    archivePrefix = "arXiv",
    primaryClass = "hep-ph",
    doi = "10.1088/1126-6708/2009/05/024",
    journal = "JHEP",
    volume = "05",
    pages = "024",
    year = "2009"
}

@article{Edsjo:1997bg,
    author = "Edsjo, Joakim and Gondolo, Paolo",
    title = "{Neutralino relic density including coannihilations}",
    eprint = "hep-ph/9704361",
    archivePrefix = "arXiv",
    reportNumber = "UUITP-11-97, MPI-PHT-97-27",
    doi = "10.1103/PhysRevD.56.1879",
    journal = "Phys. Rev. D",
    volume = "56",
    pages = "1879--1894",
    year = "1997"
}

@article{Planck:2018vyg,
    author = "Aghanim, N. and others",
    collaboration = "Planck",
    title = "{Planck 2018 results. VI. Cosmological parameters}",
    eprint = "1807.06209",
    archivePrefix = "arXiv",
    primaryClass = "astro-ph.CO",
    doi = "10.1051/0004-6361/201833910",
    journal = "Astron. Astrophys.",
    volume = "641",
    pages = "A6",
    year = "2020",
    note = "[Erratum: Astron.Astrophys. 652, C4 (2021)]"
}

@article{Biondini:2022ggt,
    author = "Biondini, Simone and Schicho, Philipp and Tenkanen, Tuomas V. I.",
    title = "{Strong electroweak phase transition in t-channel simplified dark matter models}",
    eprint = "2207.12207",
    archivePrefix = "arXiv",
    primaryClass = "hep-ph",
    reportNumber = "HIP-2022-19/TH, NORDITA 2022-050",
    doi = "10.1088/1475-7516/2022/10/044",
    journal = "JCAP",
    volume = "10",
    pages = "044",
    year = "2022"
}

@article{Bollig:2021psb,
    author = "Bollig, Julian and Vogl, Stefan",
    title = "{Impact of bound states on non-thermal dark matter production}",
    eprint = "2112.01491",
    archivePrefix = "arXiv",
    primaryClass = "hep-ph",
    doi = "10.1088/1475-7516/2022/10/031",
    journal = "JCAP",
    volume = "10",
    pages = "031",
    year = "2022"
}

@article{vonHarling:2014kha,
    author = "von Harling, Benedict and Petraki, Kalliopi",
    title = "{Bound-state formation for thermal relic dark matter and unitarity}",
    eprint = "1407.7874",
    archivePrefix = "arXiv",
    primaryClass = "hep-ph",
    reportNumber = "NIKHEF-2014-018",
    doi = "10.1088/1475-7516/2014/12/033",
    journal = "JCAP",
    volume = "12",
    pages = "033",
    year = "2014"
}

@article{Feng:2009mn,
    author = "Feng, Jonathan L. and Kaplinghat, Manoj and Tu, Huitzu and Yu, Hai-Bo",
    title = "{Hidden Charged Dark Matter}",
    eprint = "0905.3039",
    archivePrefix = "arXiv",
    primaryClass = "hep-ph",
    reportNumber = "UCI-TR-2009-06",
    doi = "10.1088/1475-7516/2009/07/004",
    journal = "JCAP",
    volume = "07",
    pages = "004",
    year = "2009"
}

@article{Cassel:2009wt,
    author = "Cassel, S.",
    title = "{Sommerfeld factor for arbitrary partial wave processes}",
    eprint = "0903.5307",
    archivePrefix = "arXiv",
    primaryClass = "hep-ph",
    reportNumber = "OUTP-0910P",
    doi = "10.1088/0954-3899/37/10/105009",
    journal = "J. Phys. G",
    volume = "37",
    pages = "105009",
    year = "2010"
}

@article{Guedes:2024vuf,
    author = "Guedes, Guilherme and Olgoso, Pablo",
    title = "{From the EFT to the UV: the complete SMEFT one-loop dictionary}",
    eprint = "2412.14253",
    archivePrefix = "arXiv",
    primaryClass = "hep-ph",
    reportNumber = "DESY-24-2",
    doi = "10.21468/SciPostPhys.20.3.074",
    journal = "SciPost Phys.",
    volume = "20",
    pages = "074",
    year = "2026"
}

@article{Gargalionis:2024jaw,
    author = "Gargalionis, John and Quevillon, Jeremie and Vuong, Pham Ngoc Hoa and You, Tevong",
    title = "{Linear Standard Model extensions in the SMEFT at one loop and Tera-Z}",
    eprint = "2412.01759",
    archivePrefix = "arXiv",
    primaryClass = "hep-ph",
    reportNumber = "DESY-24-184, ADP-24-19/T1258, KCL-PH-TH-2024-72, DESY-24-184; ADP-24-19/T1258; KCL-PH-TH-2024-72;",
    doi = "10.1007/JHEP07(2025)136",
    journal = "JHEP",
    volume = "07",
    pages = "136",
    year = "2025"
}

@article{Cepedello:2023yao,
    author = "Cepedello, Ricardo and Esser, Fabian and Hirsch, Martin and Sanz, Veronica",
    title = "{SMEFT goes dark: Dark Matter models for four-fermion operators}",
    eprint = "2302.03485",
    archivePrefix = "arXiv",
    primaryClass = "hep-ph",
    doi = "10.1007/JHEP09(2023)081",
    journal = "JHEP",
    volume = "09",
    pages = "081",
    year = "2023"
}

@article{Cepedello:2022pyx,
    author = "Cepedello, Ricardo and Esser, Fabian and Hirsch, Martin and Sanz, Veronica",
    title = "{Mapping the SMEFT to discoverable models}",
    eprint = "2207.13714",
    archivePrefix = "arXiv",
    primaryClass = "hep-ph",
    doi = "10.1007/JHEP09(2022)229",
    journal = "JHEP",
    volume = "09",
    pages = "229",
    year = "2022"
}

@article{Freitas:2014jla,
    author = "Freitas, Ayres and Westhoff, Susanne",
    title = "{Leptophilic Dark Matter in Lepton Interactions at LEP and ILC}",
    eprint = "1408.1959",
    archivePrefix = "arXiv",
    primaryClass = "hep-ph",
    reportNumber = "PITT-PACC-1404",
    doi = "10.1007/JHEP10(2014)116",
    journal = "JHEP",
    volume = "10",
    pages = "116",
    year = "2014"
}

@article{Brivio:2017vri,
    author = "Brivio, Ilaria and Trott, Michael",
    title = "{The Standard Model as an Effective Field Theory}",
    eprint = "1706.08945",
    archivePrefix = "arXiv",
    primaryClass = "hep-ph",
    doi = "10.1016/j.physrep.2018.11.002",
    journal = "Phys. Rept.",
    volume = "793",
    pages = "1--98",
    year = "2019"
}

@article{Baker:2015qna,
    author = "Baker, Michael J. and others",
    title = "{The Coannihilation Codex}",
    eprint = "1510.03434",
    archivePrefix = "arXiv",
    primaryClass = "hep-ph",
    reportNumber = "MITP-15-078",
    doi = "10.1007/JHEP12(2015)120",
    journal = "JHEP",
    volume = "12",
    pages = "120",
    year = "2015"
}

@article{Griest:1990kh,
    author = "Griest, Kim and Seckel, David",
    title = "{Three exceptions in the calculation of relic abundances}",
    reportNumber = "CFPA-TH-90-001A, BA-90-79",
    doi = "10.1103/PhysRevD.43.3191",
    journal = "Phys. Rev. D",
    volume = "43",
    pages = "3191--3203",
    year = "1991"
}

@article{Weiner:2012cb,
    author = "Weiner, Neal and Yavin, Itay",
    title = "{How Dark Are Majorana WIMPs? Signals from MiDM and Rayleigh Dark Matter}",
    eprint = "1206.2910",
    archivePrefix = "arXiv",
    primaryClass = "hep-ph",
    doi = "10.1103/PhysRevD.86.075021",
    journal = "Phys. Rev. D",
    volume = "86",
    pages = "075021",
    year = "2012"
}

@article{Gerard:1982mm,
    author = "Gerard, J. M.",
    title = "{FERMION MASS SPECTRUM IN SU(2)-L x U(1)}",
    reportNumber = "CERN-TH-3454",
    doi = "10.1007/BF01572477",
    journal = "Z. Phys. C",
    volume = "18",
    pages = "145",
    year = "1983"
}

@article{Schumann:2019eaa,
    author = "Schumann, Marc",
    title = "{Direct Detection of WIMP Dark Matter: Concepts and Status}",
    eprint = "1903.03026",
    archivePrefix = "arXiv",
    primaryClass = "astro-ph.CO",
    doi = "10.1088/1361-6471/ab2ea5",
    journal = "J. Phys. G",
    volume = "46",
    number = "10",
    pages = "103003",
    year = "2019"
}

@article{Arcadi:2017kky,
    author = "Arcadi, Giorgio and Dutra, Ma\'\i{}ra and Ghosh, Pradipta and Lindner, Manfred and Mambrini, Yann and Pierre, Mathias and Profumo, Stefano and Queiroz, Farinaldo S.",
    title = "{The waning of the WIMP? A review of models, searches, and constraints}",
    eprint = "1703.07364",
    archivePrefix = "arXiv",
    primaryClass = "hep-ph",
    doi = "10.1140/epjc/s10052-018-5662-y",
    journal = "Eur. Phys. J. C",
    volume = "78",
    number = "3",
    pages = "203",
    year = "2018"
}

@article{Cirelli:2024ssz,
    author = "Cirelli, Marco and Strumia, Alessandro and Zupan, Jure",
    title = "{Dark Matter}",
    eprint = "2406.01705",
    archivePrefix = "arXiv",
    primaryClass = "hep-ph",
    doi = "10.21468/SciPostPhysRev.1",
    month = "6",
    year = "2024"
}

@article{Belfatto:2025ids,
    author = {Belfatto, Benedetta and Blanke, Monika and Heisig, Jan and Kr{\"a}mer, Michael and Rathmann, Lena and Wilsch, Felix},
    title = "{Toward a comprehensive exploration of flavored dark matter models}",
    eprint = "2511.10490",
    archivePrefix = "arXiv",
    primaryClass = "hep-ph",
    reportNumber = "P3H-25-091, TTP25-043, TTK-25-37",
    doi = "10.1140/epjc/s10052-026-15960-1",
    journal = "Eur. Phys. J. C",
    volume = "86",
    number = "7",
    pages = "802",
    year = "2026"
}

@article{DAmbrosio:2002vsn,
    author = "D'Ambrosio, G. and Giudice, G. F. and Isidori, G. and Strumia, A.",
    title = "{Minimal flavor violation: An Effective field theory approach}",
    eprint = "hep-ph/0207036",
    archivePrefix = "arXiv",
    reportNumber = "CERN-TH-2002-147, IFUP-TH-2002-17",
    doi = "10.1016/S0550-3213(02)00836-2",
    journal = "Nucl. Phys. B",
    volume = "645",
    pages = "155--187",
    year = "2002"
}

@article{Hall:1990ac,
    author = "Hall, L. J. and Randall, Lisa",
    title = "{Weak scale effective supersymmetry}",
    reportNumber = "UCB-PTH-90/13, LBL-28879",
    doi = "10.1103/PhysRevLett.65.2939",
    journal = "Phys. Rev. Lett.",
    volume = "65",
    pages = "2939--2942",
    year = "1990"
}

@article{Chivukula:1987py,
    author = "Chivukula, R. Sekhar and Georgi, Howard",
    title = "{Composite Technicolor Standard Model}",
    reportNumber = "BUHEP-87-2, HUTP-87/A003",
    doi = "10.1016/0370-2693(87)90713-1",
    journal = "Phys. Lett. B",
    volume = "188",
    pages = "99--104",
    year = "1987"
}

@article{Biondini:2019int,
    author = "Biondini, Simone and Vogl, Stefan",
    title = "{Scalar dark matter coannihilating with a coloured fermion}",
    eprint = "1907.05766",
    archivePrefix = "arXiv",
    primaryClass = "hep-ph",
    doi = "10.1007/JHEP11(2019)147",
    journal = "JHEP",
    volume = "11",
    pages = "147",
    year = "2019"
}

@article{Kopp:2014tsa,
    author = "Kopp, Joachim and Michaels, Lisa and Smirnov, Juri",
    title = "{Loopy Constraints on Leptophilic Dark Matter and Internal Bremsstrahlung}",
    eprint = "1401.6457",
    archivePrefix = "arXiv",
    primaryClass = "hep-ph",
    doi = "10.1088/1475-7516/2014/04/022",
    journal = "JCAP",
    volume = "04",
    pages = "022",
    year = "2014"
}

@article{Brod:2017bsw,
    author = "Brod, Joachim and Gootjes-Dreesbach, Aaron and Tammaro, Michele and Zupan, Jure",
    title = "{Effective Field Theory for Dark Matter Direct Detection up to Dimension Seven}",
    eprint = "1710.10218",
    archivePrefix = "arXiv",
    primaryClass = "hep-ph",
    reportNumber = "DO-TH-17-21, DO-TH 17/21",
    doi = "10.1007/JHEP10(2018)065",
    journal = "JHEP",
    volume = "10",
    pages = "065",
    year = "2018",
    note = "[Erratum: JHEP 07, 012 (2023)]"
}

@article{Kavanagh:2018xeh,
    author = "Kavanagh, Bradley J. and Panci, Paolo and Ziegler, Robert",
    title = "{Faint Light from Dark Matter: Classifying and Constraining Dark Matter-Photon Effective Operators}",
    eprint = "1810.00033",
    archivePrefix = "arXiv",
    primaryClass = "hep-ph",
    reportNumber = "CERN-TH-2018-200",
    doi = "10.1007/JHEP04(2019)089",
    journal = "JHEP",
    volume = "04",
    pages = "089",
    year = "2019"
}

@article{Bai:2014osa,
    author = "Bai, Yang and Berger, Joshua",
    title = "{Lepton Portal Dark Matter}",
    eprint = "1402.6696",
    archivePrefix = "arXiv",
    primaryClass = "hep-ph",
    reportNumber = "SLAC-PUB-15912",
    doi = "10.1007/JHEP08(2014)153",
    journal = "JHEP",
    volume = "08",
    pages = "153",
    year = "2014"
}

@article{Brica:2026nwi,
    author = {Brica, Pavao and Elyaouti, Karim and Gabriel, Pedro and M{\"u}hlleitner, Margarete and Santos, Rui},
    title = "{Full Next-To Leading-Order Electroweak and QCD Corrections to the Relic Density in the CxSM}",
    eprint = "2607.19179",
    archivePrefix = "arXiv",
    primaryClass = "hep-ph",
    reportNumber = "KA-TP-16-2026",
    month = "7",
    year = "2026"
}

@article{Grzadkowski:2010es,
    author = "Grzadkowski, B. and Iskrzynski, M. and Misiak, M. and Rosiek, J.",
    title = "{Dimension-Six Terms in the Standard Model Lagrangian}",
    eprint = "1008.4884",
    archivePrefix = "arXiv",
    primaryClass = "hep-ph",
    reportNumber = "IFT-9-2010, TTP10-35",
    doi = "10.1007/JHEP10(2010)085",
    journal = "JHEP",
    volume = "10",
    pages = "085",
    year = "2010"
}

@article{Biondini:2026ryb,
    author = "Biondini, Simone and Greljo, Admir and Ponce D{\'\i}az, Xavier and Valenti, Alessandro",
    title = "{Charting the flavor structure of dark matter}",
    eprint = "2601.07921",
    archivePrefix = "arXiv",
    primaryClass = "hep-ph",
    doi = "10.1103/p45x-l7qb",
    journal = "Phys. Rev. D",
    volume = "113",
    number = "9",
    pages = "095021",
    year = "2026"
}

@article{Cohen:2020fcu,
    author = "Cohen, Timothy and Lu, Xiaochuan and Zhang, Zhengkang",
    title = "{Functional Prescription for EFT Matching}",
    eprint = "2011.02484",
    archivePrefix = "arXiv",
    primaryClass = "hep-ph",
    reportNumber = "CALT-TH-2020-047",
    doi = "10.1007/JHEP02(2021)228",
    journal = "JHEP",
    volume = "02",
    pages = "228",
    year = "2021"
}

@article{Ibarra:2014qma,
    author = "Ibarra, Alejandro and Toma, Takashi and Totzauer, Maximilian and Wild, Sebastian",
    title = "{Sharp Gamma-ray Spectral Features from Scalar Dark Matter Annihilations}",
    eprint = "1405.6917",
    archivePrefix = "arXiv",
    primaryClass = "hep-ph",
    reportNumber = "TUM-HEP-946-14, IPPP-14-48, DCPT-14-96",
    doi = "10.1103/PhysRevD.90.043526",
    journal = "Phys. Rev. D",
    volume = "90",
    number = "4",
    pages = "043526",
    year = "2014"
}

@article{Hahn:2000kx,
    author = "Hahn, Thomas",
    title = "{Generating Feynman diagrams and amplitudes with FeynArts 3}",
    eprint = "hep-ph/0012260",
    archivePrefix = "arXiv",
    reportNumber = "KA-TP-23-2000",
    doi = "10.1016/S0010-4655(01)00290-9",
    journal = "Comput. Phys. Commun.",
    volume = "140",
    pages = "418--431",
    year = "2001"
}

@article{Fuentes_Mart_n_2021,
   title={DsixTools 2.0: the effective field theory toolkit},
   volume={81},
   ISSN={1434-6052},
   url={http://dx.doi.org/10.1140/epjc/s10052-020-08778-y},
   DOI={10.1140/epjc/s10052-020-08778-y},
   number={2},
   journal={The European Physical Journal C},
   publisher={Springer Science and Business Media LLC},
   author={Fuentes-Martín, Javier and Ruiz-Femenía, Pedro and Vicente, Avelino and Virto, Javier},
   year={2021},
   month=feb }

@article{Di_Noi_2023,
   title={RGESolver: a C++ library to perform renormalization group evolution in the Standard Model Effective Theory},
   volume={83},
   ISSN={1434-6052},
   url={http://dx.doi.org/10.1140/epjc/s10052-023-11189-4},
   DOI={10.1140/epjc/s10052-023-11189-4},
   number={3},
   journal={The European Physical Journal C},
   publisher={Springer Science and Business Media LLC},
   author={Di Noi, Stefano and Silvestrini, Luca},
   year={2023},
   month=mar }

@article{Catena:2019gfa,
    author = "Catena, Riccardo and Emken, Timon and Spaldin, Nicola A. and Tarantino, Walter",
    title = "{Atomic responses to general dark matter-electron interactions}",
    eprint = "1912.08204",
    archivePrefix = "arXiv",
    primaryClass = "hep-ph",
    doi = "10.1103/PhysRevResearch.2.033195",
    journal = "Phys. Rev. Res.",
    volume = "2",
    number = "3",
    pages = "033195",
    year = "2020",
    note = "[Erratum: Phys.Rev.Res. 7, 019001 (2025)]"
}

@article{Electroweak:2003ram,
    author = "{The LEP Collaborations: ALEPH, DELPHI,  L3, OPAL and the  LEP Electroweak Working Group}",
    collaboration = "LEP, ALEPH, DELPHI, L3, OPAL, LEP Electroweak Working Group, SLD Electroweak Group, SLD Heavy Flavor Group",
    title = "{A Combination of preliminary electroweak measurements and constraints on the standard model}",
    eprint = "hep-ex/0312023",
    archivePrefix = "arXiv",
    reportNumber = "SLAC-R-701, LEPEWWG-2003-02, ALEPH-2003-017-PHYSICS-2003-005, DELPHI-2003-072-PHYS-937, L3-NOTE-2825, OPAL-PR-392, SLD-PHYSICS-NOTE-78, CERN-EP-2003-091, LEP-EWWG-2003-02",
    month = "12",
    year = "2003"
}

@article{CMS:2021ctt,
    author = "Sirunyan, Albert M and others",
    collaboration = "CMS",
    title = "{Search for resonant and nonresonant new phenomena in high-mass dilepton final states at $ \sqrt{s} $ = 13 TeV}",
    eprint = "2103.02708",
    archivePrefix = "arXiv",
    primaryClass = "hep-ex",
    reportNumber = "CMS-EXO-19-019, CERN-EP-2021-026",
    doi = "10.1007/JHEP07(2021)208",
    journal = "JHEP",
    volume = "07",
    pages = "208",
    year = "2021"
}

@article{ATLAS:2020zms,
    author = "Aad, Georges and others",
    collaboration = "ATLAS",
    title = "{Search for heavy Higgs bosons decaying into two tau leptons with the ATLAS detector using $pp$ collisions at $\sqrt{s}=13$ TeV}",
    eprint = "2002.12223",
    archivePrefix = "arXiv",
    primaryClass = "hep-ex",
    reportNumber = "CERN-EP-2020-014",
    doi = "10.1103/PhysRevLett.125.051801",
    journal = "Phys. Rev. Lett.",
    volume = "125",
    number = "5",
    pages = "051801",
    year = "2020"
}

@article{Garny:2015wea,
    author = "Garny, Mathias and Ibarra, Alejandro and Vogl, Stefan",
    title = "{Signatures of Majorana dark matter with t-channel mediators}",
    eprint = "1503.01500",
    archivePrefix = "arXiv",
    primaryClass = "hep-ph",
    reportNumber = "CERN-PH-TH-2015-036, TUM-HEP-985-15",
    doi = "10.1142/S0218271815300190",
    journal = "Int. J. Mod. Phys. D",
    volume = "24",
    number = "07",
    pages = "1530019",
    year = "2015"
}

@article{Arina:2020tuw,
    author = "Arina, Chiara and Fuks, Benjamin and Mantani, Luca and Mies, Hanna and Panizzi, Luca and Salko, Jakub",
    title = "{Closing in on $t$-channel simplified dark matter models}",
    eprint = "2010.07559",
    archivePrefix = "arXiv",
    primaryClass = "hep-ph",
    reportNumber = "P3H-20-058, TTK-20-35, CP3-20-47",
    doi = "10.1016/j.physletb.2020.136038",
    journal = "Phys. Lett. B",
    volume = "813",
    pages = "136038",
    year = "2021"
}

@article{Patel:2015tea,
    author = "Patel, Hiren H.",
    title = "{Package-X: A Mathematica package for the analytic calculation of one-loop integrals}",
    eprint = "1503.01469",
    archivePrefix = "arXiv",
    primaryClass = "hep-ph",
    doi = "10.1016/j.cpc.2015.08.017",
    journal = "Comput. Phys. Commun.",
    volume = "197",
    pages = "276--290",
    year = "2015"
}

@article{Giacchino:2013bta,
    author = "Giacchino, Federica and Lopez-Honorez, Laura and Tytgat, Michel H. G.",
    title = "{Scalar Dark Matter Models with Significant Internal Bremsstrahlung}",
    eprint = "1307.6480",
    archivePrefix = "arXiv",
    primaryClass = "hep-ph",
    reportNumber = "ULB-TH-13-09",
    doi = "10.1088/1475-7516/2013/10/025",
    journal = "JCAP",
    volume = "10",
    pages = "025",
    year = "2013"
}

@article{ATLAS:2019lff,
    author = "Aad, Georges and others",
    collaboration = "ATLAS",
    title = "{Search for electroweak production of charginos and sleptons decaying into final states with two leptons and missing transverse momentum in $\sqrt{s}=13$ TeV $pp$ collisions using the ATLAS detector}",
    eprint = "1908.08215",
    archivePrefix = "arXiv",
    primaryClass = "hep-ex",
    reportNumber = "CERN-EP-2019-106",
    doi = "10.1140/epjc/s10052-019-7594-6",
    journal = "Eur. Phys. J. C",
    volume = "80",
    number = "2",
    pages = "123",
    year = "2020"
}
\end{document}